\documentclass[a4paper, 12pt]{amsart}

\usepackage[utf8]{inputenx}
\usepackage[english]{babel}

\usepackage{enumerate}
\usepackage{amsthm}
\usepackage{amsmath}
\usepackage{amssymb}
\usepackage{scalerel}
\usepackage{mathtools}
\usepackage{bbm}
\usepackage{cases}
\usepackage{tikz}
\usepackage{IEEEtrantools}
\usepackage{hyperref}
\usepackage{graphicx}
\usepackage{emptypage}
\usepackage{color}
\usepackage{thmtools}
\usepackage{thm-restate}
\usepackage{todonotes}

\usepackage{comment}

\newtheorem*{thm*}{Theorem}
\newtheorem{thm}{Theorem}[section]

\newtheorem{lemma}[thm]{Lemma}
\newtheorem{prop}[thm]{Proposition}
\newtheorem{corollary}[thm]{Corollary}

\theoremstyle{definition}

\newtheorem{remark}[thm]{Remark}

\newtheorem*{remark*}{Remark}

\numberwithin{equation}{section}

\newcommand{\R}{\mathbb{R}}
\newcommand{\C}{\mathbb{C}}
\newcommand{\Z}{\mathbb{Z}}
\newcommand{\e}{\ensuremath{\varepsilon}}

\newcommand{\med}{\medskip\noindent}
\newcommand{\Gconv}{\xrightarrow{\Gamma}}

\newcommand{\bra}[1]{\left\langle {#1} \right\rangle}
\newcommand{\res}{\mathbin{\vrule height 1.6ex depth 0pt width 0.13ex\vrule height 0.13ex depth 0pt width 1.3ex}}

\DeclareMathOperator{\supp}{supp}
\DeclareMathOperator{\cl}{cl}

\usepackage{tikz}
\usetikzlibrary{calc} 
\usepackage[utf8]{inputenx}
\usepackage[english]{babel}
\usepackage{todonotes}

\colorlet{darkred}{red!80!black}
\colorlet{darkblue}{blue!80!black}
\colorlet{darkgreen}{green!60!black}

\def\e{\varepsilon}

\def\R{\mathbb{R}}
\def\N{\mathbb{N}}
\def \O {\Omega}
\def\Ob {\ov\O}

\def\d{\delta}
\def\u{{\overline u}}

\newcommand{\one}{ 1\!\!\!1}
\def\1{{{\bf 1}}}

\def\GG{\mathcal{G}}

\def\EE{\mathcal{\xi}}
\def\PP{\mathcal{P}}
\def\CC{\mathcal{C}}
\def\f{\varphi}
\def\a{\alpha}

\def\f{\varphi}
\def\la{\lambda}

\def\ds{\displaystyle}
\def\eps{{\varepsilon}}
\def\N{\mathbb{N}}
\def\R{\mathbb{R}}
\def\Z{\mathbb{Z}}
\def\O{\Omega}

\def\B{\mathcal{B}}
\def\C{\mathcal{C}}
\def\E{\mathcal{E}}
\def\FF{\mathcal{F}}

\def\MM{\mathcal{M}}
\def\I{\mathcal{I}}

\def\LL{\mathcal{L}}
\def\M{\mathcal{M}}
\def\PP{\mathcal{P}}
\def\FF{\mathcal{F}}

\def\SS{\mathcal{S}}

\newcommand{\be}{\begin{equation}}
\newcommand{\ee}{\end{equation}}

\newcommand{\weak}{\stackrel{*}{\rightharpoonup}}

\newcommand{\ov}{\overline}

\DeclareMathOperator{\spt}{spt}

\title[Mean field theory for short-range interaction functionals] {Mean field theory for a general class of short-range interaction functionals}

\begin{document}

\date{\today}
\author{Guy Bouchitt\'e}
\address{ Imath \\ Universit\'e de Toulon, BP 20132\\  83957 La Garde Cedex- FRANCE}
\email{bouchitte@univ-tln.fr}
\thanks{ Part of this work was prepared at FCFM, Univ. Concepci'{o}n. The first author warmly thanks this institution for its financial support and hospitality during the fall of 2022.}  

%
%% Repeat the preceding commands for additional authors, commenting out lines
%% which should not appear
%% If an author has an ORCID, this should be added as shown below
\author{Rajesh Mahadevan}
\address{Depto. de Matem\'{a}tica, FCFM, Univ. de Concepci\'{o}n, Concepci\'{o}n, Chile}
\email[Mahadevan]{rmahadevan@udec.cl}
\thanks{The second author acknowledges the support of Dirección de Postgrado, UdeC for the funding received through the project UCO 1866 }
%% Each \address command increments a counter. If you want to refer to an address already
%%  listed for a previous author, use the command below in lieu of \address. 
%% The number is the appearance order of that address (among all addresses)
% \addressSameAs{1}{<repeat address 1>}
% This can be done multiple times
%\addressSameAs{1}{Road to the 5th problem avenue, Germany}
%\addressSameAs{2}{Center for experimental machines, United Kingdom}

%% The grant number can be inserted in the database
%% This won't be printed. It should be acknowledged in \thanks as above.
% \CDRGrant[UKRC]{2019-$$55900}

%% If you want to inform the reader of the paper about your
%% supplementary material, you can refer to it this way. The file
%% itself should be placed in a directory called Attach.

%\ESM{Supplementary material for this article is supplied as a separate 
%archive available from  the journal's
%website under 
%%% this will be replaced by the DOI of the article
%article's URL 
%%\printDOI\ 
%or from the author.}

%% If yo have supplementary material, you have to declare it this way (the
%% file will be copied and linked on the website
%% PDF is the default file type
% \CDRsupplementaryTwotypes{supplementary-material}{\cdrattach{supplement-doc.pdf}}
%% For another file type you should declare the mime-type
% \CDRsupplementaryTwotypes[application/zip]{supplementary-material}{\cdrattach{mycode.zip}}

%% Mathematical classification (2020)
%% For previous classifications, use \subjclass[2010]{00X99}

%% Abstract should be placed before \maketitle (and, in fact, before
%% \begin{document is best)
\begin{abstract} 
   In  models of $N$  interacting particles in $\R^d$ 
	as in Density Functional Theory or  crowd motion, the repulsive cost is usually described by a two-point function 
	$c_\e(x,y) =\ell\Big(\frac{|x-y|}{\e}\Big)$ where $\ell: \R_+ \to [0,\infty]$ is decreasing to zero at infinity and parameter $\e>0$ scales the interaction distance. In this paper we identify the mean-field energy of such a model in the short-range regime $\e\ll 1$ under the sole assumption  that $\exists r_0>0 \ : \ \int_{r_0}^\infty \ell(r) r^{d-1}\, dr <+\infty$.
	This extends recent results \cite{hardin2021, HardSerfLebl, Lewin} obtained in the homogeneous case $\ell(r) = r^{-s}$ where $s>d$.

\end{abstract}

%%% Abstract in the other language
%\begin{altabstract} 
%Dans les mod\`eles d'interactions \`a $N$ particules dans $\R^d$ qui interviennent en th\'eorie de la fonctionnelle de densit\'e ou dans l'\'etude des  mouvements de foule,  le c\^out  r\'epulsif est en g\'en\'eral d\'ecrit par une fonction \`a deux points du type $c_\e(x,y) =\ell\Big(\frac{|x-y|}{\e}\Big)$ o\`u $\ell: \R_+ \to [0,\infty]$ tend vers  z\'ero \`a l'infini en d\'ecroissant et o\`u   $\e>0$ est un param\`etre d' \'echelle pour la distance d'interaction. Dans cette note, nous identifions l'\'energie  limite de champ moyen 
%d'un tel mod\`ele dans le r\'egime \`a courte port\'ee $\e\ll 1$, sous la seule hypoth\`ese d'int\'egrabilit\'e \`a l'infini  $\int_{r_0}^\infty \ell(r) r^{d-1}\, dr <+\infty$. Nous \'etendons ainsi des r\'esultats r\'ecents \cite{HardSerfLebl,hardin2021, Lewin} obtenus dans le cas homog\`ene $\ell(r) = r^{-s}$ o\`u $s>d$.
%%  Ce document est un petit guide d'utilisation de la classe \LaTeX
%%  pour les articles de \emph{\currentjournaltitle}.
%\end{altabstract}
\maketitle



% Use the \maketitle command after the abstract
\maketitle

% Example of section
\section{Introduction}

We consider a repulsive interaction function on $(\R^d)^N$ of the kind
\begin{equation}\label{cNeps}
c^\e_N(x_1\,\dots,x_N)=  \sum_{i\not=j} \ell \Big(\frac{|x_i-x_j|}{\e}\Big).
\end{equation}

where:
\begin{itemize}
\item $N$ is the number of particles in $\R^d$;  

\smallskip
\item $\e>0$  scales the interaction distance between particles.

\smallskip
\item the two-particle cost $\ell: [0,+\infty]\to [0+\infty]$ satisfies:
\end{itemize}

\smallskip
\begin{itemize}
\item [(H1)]  $\ell$ is l.s.c. and $\ell(0)>0$  ( $\ell(0)=+\infty$ is allowed)

\med
\item [(H2)]  $\exists r_0\ge 0$ such that $\ell$ is finite and non increasing on $[r_0,+\infty)$ and $\ds \lim_{r\to \infty} \ell(r) = 0$.
\end{itemize}
%\end{itemize}>0$ suc

In the whole paper, we denote by $\Ob$  the closure of a smooth domain $\O\subset\R^d$ where the  $N$ particles are located.
 We speak of a \emph{confined system} when $\Ob$ is compact (container).   
 Given a continuous exterior potential $U: \Ob\to \R$,  we consider the finite dimensional problem:
 \begin{equation}\label{infN}
\mathcal{\E}_N^\e(\O,U) :=  \inf \left\{ h_N\ c_N^\e(x_1\,\dots,x_N) + \frac1{N} \sum_{i=1}^N U(x_i): x_i \in \Ob\right\} ,
\end{equation}
where $h_N$ is a suitable chosen normalization factor. 
Since the seminal work of Choquet in 1958  \cite{choquet1958diametre} and the growing interest of the quantum and statistical mechanics community,
a lot of work has been  devoted to the limit behavior of $\mathcal{\E}_N^\e(\O,U)$ as $N\to\infty$ ($\e$ fixed)
as well as the characterization of the weak cluster points of the empirical measures associated with $N$-point configurations of minimal energy. The cornerstone  of the mean field theory consists in  identifying a limit energy functional on measures whose minimizers are precisely  these cluster points. 

\subsection{State of the art}\   The scaling factor $h_N$  in \eqref{infN}  must be selected so that the limit of the infimum
belongs to $(0,+\infty)$.   
In turn  this issue  relies heavily  on the integrability properties of the function $g(x)= \ell(|x|)$.  
Let us report on two cases of major interest:

\subsubsection{Long range interaction case} \  Here $\O=\R^d$ and we take $\e=1$. Moreover in addition to (H1)(H2), we assume  that $g\in L^1_{loc} (\R^d)$ i.e.:
\begin{equation}\label{L1loc}
\int_0 ^1 r^{d-1} \ \ell(r) \, dr <+\infty.
\end{equation}
   
In that case, a relevant choice is  $h_N= \frac1{N^2}$  meaning roughly that  the interaction energy  is averaged over all  pairs  of  distinct points in $\{x_1,x_2,\dots x_N\}$. 
The identification of the mean-field energy is well known in the case of Riesz  potentials  $\ell(r) = \frac1{r^{s}}$
for $0<s<d$,  in the Logarithmic case  $\ell(r)=- \log(r)$  for $d=2$ and more generally for $\ell$ of positive type i.e. such that
the Fourier transform of $g(x)=\ell(|x|)$ is positive in $\R^d$ (see for instance the monograph by S. Serfaty \cite{Serfaty2015}). 
It is given by a  non-local functional, the so called \emph{Direct energy}:  
\begin{equation}\label{def:Dell}
D_\ell(\rho) :=  \iint \ell(|x-y|)  \, \rho\otimes\rho(dxdy) .
\end{equation}
Accordingly the limit problem associed with \eqref{infN} reads:
$$\mathcal{\E}_\infty (\O,U) =  \inf  \left\{  D_\ell(\rho) + \int U \, d\rho  \ :\  \rho\in \PP(\R^d) \right\},  $$
where the infimum is reached at a unique configuration provided $U$ growths suitably at infinity. 
 At this stage, a few comments are in order:

\med
-  in the case of a confining external potential $U$, there are several impressive works devoted to the next order asymptotics  \cite{Serfaty2015, petrache2017next, serfaty2018systems,cotar2019next}  in the case of Riesz potentials $\ell(r)= r^{-s}$  for $d\ge 3$ and $d-2\le s< d$ revealing  an asymptotic behavior as $N\to \infty$ of the form: 
\begin{equation}
\lim_{N\to\infty}  N^{1-\frac{s}{d}} \left(\mathcal{\E}_N(\O,U) - \mathcal{\E}_\infty (\O,U)\right) \ =\  C(s,d)\, \int (\rho_U)^{1+\frac{s}{d}}  ,  
\end{equation}
 where $\rho_U$ is the unique minimizer realizing $\mathcal{\E}_\infty (\O,U)$.
 
 \med
 - if $U$ remains bounded at infinity  (for instance a Coulomb potential    vanishing at infinity), the existence of  a minimizer $\rho_U$  may fail due to a loss of mass at infinity along minimizing sequences. A relaxation procedure leads to consider minimizers in the class of sub-probabilities $\rho\in \PP_-(\R^d)$ and involves  the weak* lower semicontinuous convexification of the Direct energy $D$.
 If $\ell$ is of positive type, this relaxed energy coincides with the natural 2-homogeneous extension of $D$ to $\PP_-(\R^d)$
 while almost nothing is known if $\ell$ is merely locally integrable. For further details and examples of relaxed minimizers, we refer to the recent paper  \cite{BindBou2022}.

 \subsubsection{Short range interaction case} \  Following an idea developed for the hard spheres model   \cite{BindBou2022}, we  look now at $\e$ as a small parameter tending to zero with a prescribed speed as $N\to \infty$. If one thinks to a container $\Ob$ of unit volume and $\e$ to be
  the average distance of a particle to the others,  it is natural to  consider an asymptotic analysis where the
    product $\e^d \, N$ remains constant or converges to a given intensity factor $\kappa\in (0,+\infty)$. 
    In a crowd model, this factor $\kappa$ is related to a congestion  ratio (see the hard speres model in Section 3.4 and Remark \ref{filling}).
    This of course means that we need to assume that $\e\sim N^{-\frac1{d}}$. Accordingly, in order to obtain a precise scaling for $h_N$ 
    ensuring a non-trivial behavior of the infimum  \eqref{infN}, it is crucial to make an additional integrability  on $\ell$ at infinity namely
    $$   \int_{r_0}^{+\infty}   \ell(r) r^{d-1} dr <+\infty  .\leqno(H3)$$
  It turns out that, under $(H3)$,  the right scaling factor in  \eqref{infN} is $h_N=\frac1{N}$ in contrast with the long range case. This covers the case of  hyper singular Riesz potentials $\ell (r) =r^{-s}$ with $s>d$. For such potentials the parameter 
  $\e$ can be dropped thanks to the homogeneity and the normalized interaction energy becomes  
  $\frac1{N^{1+ \frac{s}{d}}}\, c_N^{1}(x_1\,\dots,x_N)$. Under the latter scaling, it was proved recently \cite{hardin2021, HardSerfLebl} that the mean field energy is a local funtional defined on absolutely continuous measures $\rho=u \,\LL^d\res\O$  by  $F(\rho) = C(s,d) \, \int_\O u^{1+ \frac{s}{d}}\, d\LL^d,$ being $C(s,d)$  a universal constant. 
  However,  extending this result to more general costs seems to be difficult in the framework developed in \cite{hardin2021},
except possibly if $\ell$ is very close to a power potential. 

\subsection{Our contribution}\ This paper proposes a significant simplification of asymptotic analysis in the short range case. The approach is based on two
components: first, treating the \emph{interaction distance} $\e$ in \eqref{cNeps} as an infinitesimal parameter, and second, using an $\epsilon$-counterpart of the traditional empirical 
measure  frequently utilized in mean-field theory. 

 Thus, for every cost $\ell$ satisfying $(H3)$, we can determine the mean field energy in terms of a \emph{local} integral functional of the type $\int_\Omega f_\ell(u)\, dx$, where $u= \dfrac{d\rho}{dx}$ denotes the local particle density and $f_\ell$ is a convex integrand that exhibits super-linear growth at infinity. This expands upon previous findings \cite{BindBou2022} that were limited to the hard-spheres model (where $\ell(r)= 
 +\infty $ when $r<1$ and $\ell(r)=0 $ otherwise). Similarly, this permits to handle the case of hyper-singular Riesz potentials  $\ell(r) = r^{-s}$ for $s>d$ analyzed in \cite{hardin2021, HardSerfLebl}.  
  
It is noteworthy that fulfilling the integrability condition $(H3)$ is crucial and cannot be sidestepped. 
 When assuming that $\O$ is bounded, a cost $\ell$ that satisfies $\int_{r_0}^{+\infty} r^{d-1} \ell(r) \, dr=+\infty$ would give rise to an infinite limit in \eqref{infN} if the scaling is by $h_N=\frac1{N}$ and $\e^d \, N \sim 1$ (see Remark \ref{lim=infty}). 
  
\subsection{Setting of the asymptotic problem and notations.}

%\setlength{\leftmargini}{6pt}
%\begin{enumerate}
%
From now on, $\O$ will be a \emph{bounded} domain of $\R^d$ with  Lipschitz boundary ($\partial \O$ needs to be $\LL^d$- negligible) and 
 we consider a cost function $\ell$ which satisfies the standing assumptions $(H1), (H2)$ and $(H3)$.

\medskip
For purposes of presentation, we utilize the infinitesimal length $\e$ as the main parameter while the number of particles $N=N_\e$ approaches infinity as $\e$ tends towards zero, following the scale $N_\e \sim \kappa\, \e^{-d}$ where $\kappa$ is a positive constant.
Later, we will establish that assigning a value to $\kappa$ is not required because a uniform bound on the $N_\e$-point interaction energy 
 will automatically result in $\limsup_{\e\to 0} N_\e\, \e^d <+\infty$. 
\medskip
We can now incorporate measures in $\Ob$ into a variational framework for addressing the mean field problem.
For each finite subset $S\subset \Ob$, we define its $\e$-scaled empirical measure as follows:
\begin{equation}\label{eps-muS}  
\rho_S^\e \ :=\ \e^d \, \sum_{x\in S} \, \delta_x .
\end{equation}
This measure belongs to the set of non-negative Borel measures on $\Ob$, denoted by $\MM_+(\Ob)$. Here, $\|\rho\|$ represents the total mass, which may be infinite, of any element $\rho\in \MM_+(\Ob)$. Through this, we can observe that $\|\rho_S^\e \|\,=\, \e^d \, \sharp(S)$, which could deviate from the classical empirical measure of $S_\e$ with a total mass equal to $1$.  
A  key avantage of this approach is that it allows to avoid  the non-local constraint that all competitors must belong to the subclass $\PP(\Ob)$ of probability measures.

\med
Next, we define the $\eps$-scaled interaction energy of a discrete set  $S\subset \R^d$ as follows:
\begin{equation}\label{def:EEeps}
\EE_{\ell,\e}(S)=   \sum_{(x,y)\in S^2\setminus \Delta} \ell\left(\frac{|x-y|}{\e}\right)
\quad \text{where}\quad  \Delta:=\{(x,x) : x\in \R^d\}.
\end{equation}
We will refer to the interaction energy corresponding to $\e=1$ as the ``ground interaction energy'', denoted by $\EE_\ell$. When the cost function $\ell$ is fixed, we will 
use $\EE_\e$ instead of $\EE_{\ell,\e}$. 
With this in mind, we can define a scaled energy functional $F_\e:\MM_+(\Ob)\rightarrow [0,+\infty]$ for every $\e>0$ in the following way:
\begin{align} \label{def:Feps}
F_\e(\rho) =
\begin{cases}  \e^d\, \EE_\e(S) & \text{if  $\exists S\subset \Ob$
such that $\rho= \rho^\e_S$}\\
 +\infty &\text{otherwise.}  \end{cases}\end{align}

The discrete problem  \eqref{infN} for $h_N=\frac1{N}$ and $N\, \e^d\sim \kappa$ 
can then be expressed through the relation:
$$ \kappa\ \mathcal{\E}_N^\e(\O,U) \sim \  \inf \left\{ F_\e(\rho) +   \,  \int_{\Ob} U \, d\rho \ :\ \rho\in \MM_+(\Ob)\right\}\quad\text{as}\quad \e\to 0 . $$
Accordingly  the mean-field energy will be represented by a functional $F:\MM_+(\Ob)\to [0,+\infty]$ characterized by the property that, for every $U\in \C(\Ob)$, one has the convergence of infima
$$  \inf \left\{ F_\e(\rho) +   \,  \int_{\Ob} U \, d\rho \right\} \to \inf \left\{ F(\rho) +   \,  \int_{\Ob} U \, d\rho \right\} $$
accompanied by the tight convergence of minimizers. This falls squarely within the $\Gamma$-convergence theory   (\cite{attouch1984, dal2012, braides2002gamma}) on which we rely to support our results.
 
 \medskip  
The paper is organized as follows: in Section 2, we establish a lower bound  which  allows to  obtain the strong equi-coercivity  of the sequence $(F_\e)$; in addition we show that any weak* cluster point of a sequence $(\rho_\e)$ with  uniformly bounded energy is absolutely continuous with respect to the Lebesgue measure; in Section 3, we state the $\Gamma$-convergence of $F_\e$ as $\e\to 0$ to  a convex functional of the form
 $F(\rho)=\int_\O f_\ell(\frac{d\rho}{dx}) \, dx$,  where the effective integrand $f_\ell$  growths  at least  quadratically  at infinity. It is given by 
 the thermodynamical limit of a subadditive set funtion (\emph{Krengel's} theorem). 
 Some examples and applications are given. 
 The Section 4 is devoted to the proof of the main theorem.

%\red{We conclude this introduction with the hope that, by adapting some ideas developed in the present paper, it will possible
%to derive a general framework for tackling the asymptotic of next order in the long range setting and for non homogeneous costs $\ell$. }

 \subsection*{Notations:}

\begin{enumerate}

\item [-] $B(x,r)$ is the open 
ball of the Euclidean space $\R^d$ centered at $x$ and of radius $r$ ; if $x=0$, we simply denote $B_r$;

\item [-] $Q_k$ denotes the hypercube $[- k/2, k/2)^d$, $Q (x_0, r):= x_0 + r\, Q_1$;

\item [-] $\Delta = \{(x,x) : x\in\R^d\}$ stands for the diagonal of $\R^d$; 

\item [-]  $\sharp(S)$ denotes the counting measure of a subset $S\subset \R^d$  ($+\infty$ if $S$ is infinite);

\item [-]  $\LL^d$ is the  Lebesgue measure in $\R^d$; given any Borel set $B$, $|B|$ is a short notation  for $\LL^d(B)$ ;  $\omega_d$ is such that  $|B(x,r)|= \omega_d   r^d$;

\item [-]  $\CC(\Ob)$ denotes the Banach space of continuous functions on the compact subset $\Ob$ equipped with the uniform norm;
 
\item [-] $\MM(\Ob)$  stands for the space of signed Radon measures on $\Ob$ equipped with the total variation norm;

\item [-]   $\PP_-(\Ob)$ (resp. $\PP(\Ob)$) is the subset of Borel measures $\mu\in \MM_+(\Ob)$ such that $\|\mu\| := \mu(\Ob) \leq 1$ (resp. $\|\mu\| = 1$).

%\item[-] Given $\mu \in \MM_+(\Ob)$ and $h \in \R^d$,  $\tau_h \mu$  the translation of $\mu$ by the vector $h$
%is the element of $\MM_+(\Ob+h) $ defined by   $\tau_h \mu(E) = \mu(E-h)$ for every Borel set $E\subset \Ob+h$.

\item [-]  The topological support of $\mu \in \MM_+(\Ob)$ is denoted  $\supp(\mu)$ 
while $\mu\res A$ represents its restriction to a Borel subset $A\subset \Ob$;   

\item [-] The bracket $\bra{\cdot, \cdot}$ will denote the duality between $\CC(\Ob)$ and $\MM(\Ob)$:
\[ \bra{v, \mu} = \int v d\mu , \]
This duality induces  the weak* topology on $\MM(\Ob)$ which can be identified with the dual of  $\CC(\Ob)$; as $\Ob$ is compact, the weak* convergence $\mu_h \weak \mu$ in  $\MM_+(\Ob)$ implies the tight convergence since $\|\mu_h\| = \bra{1,\mu_h} \to \bra{1,\mu}= \|\mu\|$.

\item [-] To any non-empty  set $A$, we associate the functions: 
$$\one_A(x)=\begin{cases} 1 & \text{if $x\in A$}\\
0  & \text{otherwise}\end{cases}\quad, \quad \chi_A(x)=\begin{cases} 0 & \text{if $x\in A$}\\
+\infty  & \text{otherwise}\end{cases} .$$

%\item[-]  $\PP(\PP_-(\R^d))$ denotes the set of Borel probabilities measures on $\PP_-(\R^d)$ (seen as 
%a weakly* compact metrizable space).

\end{enumerate}

\medskip

\section{Energy estimates and compactness.}

\medskip
We begin with some elementary properties of the set function $\EE_\ell(S)$ (defined by \eqref{def:EEeps} for $\e=1$).
\begin{lemma} \label{subp}  Let $S_1, S_2$  be finite disjoint subsets of  $\R^d$.
Then we have:

\med (i)\ \emph{(super-additivity)}
\begin{equation*}
\EE_\ell (S_1 \cup S_2)\  \ge \ \EE_\ell (S_1) + \EE_\ell (S_2) . 
\end{equation*}

\med (i)\ \emph{(sub-additivity at large distance)}
\begin{equation*}
\EE_\ell (S_1 \cup S_2)\  \le \ \EE_\ell (S_1) + \EE_\ell (S_2)
+  2\, \ell_+ (\eta )  \  \sharp(S_1)\, \sharp(S_2)   , 
\end{equation*}
where $\eta:= {\rm dist}(S_1,S_2)$.
\end{lemma}
\begin{proof}   \  Since $S_1$ and $S_2$ are non-intersecting, we can split $(S_1\cup S_2)^2$ in four disjoint pieces as follows:
$$(S_1\cup S_2)^2  = (S_1\times S_1) \cup (S_2\times S_2) \cup (S_1\times S_2) \cup (S_2\times S_1) .$$
The  inequality  (i) is then straightforward whereas, for the (ii),
we simply majorize by $\ell_+(\eta)$ the contribution  $\ell(|x-y|)$  of each pair $(x,y)$ in $(S_1\times S_2) \cup (S_2\times S_1)$  where $|x-y|\ge \eta$ holds.
 \end{proof}
Recalling the notations  \eqref{eps-muS} and \eqref{def:Feps} given in the introduction, we may rewrite  a $\e$- rescaled version of Lemma \ref{subp} as follows: for every $\e>0$ and all pairs of non-intersecting subsets $(S_\e', S_\e'')$, one has
  \begin{equation}\label{subp-eps}
F_\e (\rho_\e') + F_\e (\rho_\e'') \ \le \  F_\e (\rho_\e' +\rho_\e'')  \ \le \ F_\e (\rho_\e') + F_\e (\rho_\e'') 
+   \frac{ 2 \ell_+ (\eta_\e )}{\e^d}  \  \| \rho_\e'\| \| \rho_\e''\| ,
\end{equation}
where $\rho_\e':=  \rho_{S_\e'}^\e$, $\rho_\e'':=  \rho_{S_\e''}^\e$ and $\eta_\e= \e^{-1}\, {\rm dist}(S_\e',S_\e'')$ .

\subsection{A fundamental lower-bound}
 Since we are only assuming that $\ell(r)$ is decreasing for suitably large $r$, we need to define :
\begin{align} \label{ell-pm} \ell_+(r) := \sup\left\{ \ell(s) : s\ge r \right\} \quad,\quad 
\ell_-(r) :=\inf_{x,y \in [0,r]^d }  \ell(|x-y|) 
\end{align}
which are   monotone non-increasing  and satisfy: 
$$ \ell_+(r)=\ell(r) \quad \forall r\ge r_0 \quad,\quad  \ell_-(r)\ =\ \inf \left\{ \ell(s) : s\le  r\sqrt{d}  \right\}\ \le\ \ell_+(r\sqrt{d}).$$

\medskip
Next, in the same line as in the survey \cite{Lewin}, we derive  a very simple but fundamental lower bound for $\EE_\ell(S)$  when $S$ is a $N$-point system contained in a  Borel subset  $B$ of finite volume in $\R^d$. For such a $B$ and any $\d>0$, we denote  
by $ m_\d(B)$  the minimal number of disjoint $\d$-hypercubes 
  $Q_j= x_j+ [-\d/2,\d/2[^d$ such that $\ds B\subset \cup_{j=1}^{m_\d(B)} \ov{Q_j}$.
  Then it's easy to check that $ m_\d(B)\sim \d^{-d}\, \LL^d(B)$
as $\d\to 0$. For such subsets $B$, we will often use the following equivalent  version obtained by keeping $\d$ fixed while using large dilations:
\begin{equation}\label{mdelta}
\lim_{\e\to 0}  \, \e^d \, m_\d \Big( \frac{B}{\e}\Big)  \ =\  \d^{-d} \, \LL^d(B).
\end{equation}

\begin{lemma} \label{lemma-low} Let $B\subset \R^d$ be a Borel subset such that $\LL^d(B)<+\infty$.   
 Then for any $N$-point system $S\subset B$ and any $\d>0$, we have
\begin{equation}\label{fundlower}
\EE_\ell(S) \ge \  N\, \ell_-(\d)\, (\zeta-1)_+ 
 \quad \text{where}\quad \zeta= \frac{N}{m_\d(B)}.
\end{equation}
 
\end{lemma}

\begin{proof}

Let $\{ Q_j, \, 1\le j\le m_\d(B)\}$ be a covering of $B$ by disjoint $\d$-hypercubes $Q_j$ and 
denote $n_j= \sharp(S\cap Q_j)$. We have $N = \sum n_j$ while, by the super additivity of $\EE_\ell$ (see Lemma \ref{subp})  and the definition of $\ell_-(\d)$, we have:
$$ \EE_\ell(S) \ge \sum_{j=1}^{m_\d(B)} \EE_\ell(S\cap Q_j)\ \ge\  \ell_-(\d)\, \sum_{j=1}^{m_\d(B)} n_j(n_j-1) .$$
If $N > m_\d(B)$ (i.e. $\zeta > 1$), the desired lower bound \eqref{fundlower} follows by noticing that the infimum
$$ \inf \left\{ \sum_{j=1}^{m_\d(B)} t_j(t_j-1)\ :\ \sum_{j=1}^{m_\d(B)} t_j= N,\ t_j \in \R \right\} $$
is reached for $t_j = \frac{N}{m_\d(B)}=\zeta \ ,\forall j .$ 

If $\zeta\le1$  the inequality \eqref{fundlower} is trivial since $\EE_\ell(S)\ge 0$.
\end{proof}

\begin{remark} By the lower semicontinuity assumption (H1), 
since $\ell_-(0)=\ell(0) \in (0,+\infty]$, we may always find a $\d>0$ such that $\ell_-(\d) >0$.
Note that the inequality \eqref{fundlower} is still valid if $\ell_-(\d) =+\infty$ provided we agree that
 $0\times +\infty =0$  (this situation occurs when $\ell=+\infty$ on an interval $[0,\d_0)$).

\end{remark}  
%%%%%%%%%%%%%%%%%
%%%%%%%%%

\subsection{Strong coercivity and compactness}

We recall the definition of the functional $F_\e$ given  in \eqref{def:Feps}  whose domain consists of $\e$-empirical measures of discrete subsets $S_\e\subset\Ob$ (see the definition \eqref{eps-muS}). 

\begin{lemma}[strong coercivity]\label{coercivity} Assuming that $\ell$ satisfies (H1) (H2), let $\d>0$ be such that $\ell_-(\d) \in (0,+\infty]$
and set $\a := \dfrac{\d^d \, \ell_-(\d)}{\LL^d(\O)}.$ 

 Then we have: 
\begin{equation}\label{strong-coercice}
\liminf_{\e\to 0} \frac{F_\e (\rho_\e)}{ \| \rho_\e\|^2} \ \ge\ \alpha \ , 
\end{equation}
 whenever $(\rho_\e)$ is a sequence such that   $\|\rho_\e\| \to +\infty$. 
\end{lemma}
\begin{proof}\ 
 Without loss of generality, we may assume that $F_\e (\rho_\e) <+\infty$. Thus $\rho_\e= \rho^\e_{S_\e}$ for a suitable $N_\e$-point system
  $S_\e\subset\Ob$ while:
$$   F_\e(\rho) =  \e^d \EE_\e(S_\e) = \e^d\, \EE_\ell  \left( \dfrac{S_\e}{\e}\right)\quad,\quad  \|\rho_\e\| = N_\e\,  \e^d . $$
By applying the lower bound  \eqref{fundlower} to the subset $\e^{-1} S \subset \e^{-1} \Ob$, we get:
\begin{equation}\label{ineq-coercive}
F_\e(\rho_\e)\, \ge\,  \ell_-(\d) \, \|\rho_\e\| \, \left( \frac{\|\rho_\e\|}{ \beta_\e} - 1\right)_+\quad  \text{where}\  \beta_\e =\e^d \, m_\d\left( \frac{\Ob}{\e}\right). 
\end{equation}
From \eqref{mdelta}, we know that  $\beta_\e\to  \d^{-d}\, \LL^d(\O)$. Hence \eqref{strong-coercice} follows by dividing \eqref{ineq-coercive}
by $\|\rho_\e\|^2$ and passing to the limit as $\e\to 0$. 
   \end{proof}

\medskip

\begin{prop}[compactness]\label{compactness-Feps}  Assuming that $\ell$ satisfies (H1) and (H2),  let $U:\Ob\to \R$ be a bounded Borel function 
and  $(\rho_\e)$  a sequence in $\MM_+(\Ob)$   such that 
\begin{equation}\label{finite-energy}
\sup_{\e}  \left( F_{\e}(\rho_\e) + \int U\, d\rho_\e \right)  <+\infty .
\end{equation}

Then :
 
\med 
 (i)\ there exists a constant $C$ such that 
 $$   \|\rho_\e \| + F_\e(\rho_\e)\ \le\ C <+\infty\quad \text{for every $\e>0$} .$$
 (ii)\  any weak* cluster point of $(\rho_\e)$ is of the form $\rho = u \, \LL^d \res \O$ with $u\in L^1(\O)$.

\end{prop} 

\begin{remark} The proposition above implies  that sequences of point configurations $S_\e$ with equi-bounded energies 
admit a finite limiting intensity factor  $\kappa:= \limsup_\e \e^d \sharp(S_\e)$  while, by the assertion (ii),
$S_\e$ is  not allowed to concentrate anywhere as $\e\to 0$.
\end{remark}

\med

 \begin{proof}  Suppose that $\|\rho_\e\|$ has no upper bound. Then   
the  uniform energy upper bound \eqref{finite-energy} implies that
 $$\liminf_{\e\to 0} \dfrac{F_\e(\rho_\e)}{\|\rho_\e\|} \ \le\ \sup_{\Ob} |U| ,$$
while by \eqref{strong-coercice}  the left hand member of the previous inequality 
is infinite. So there is a contradiction and we can conclude that $(\rho_\e)$ is bounded. Then it follows from  \eqref{finite-energy} that  $F_\e(\rho_\e)$ is bounded as well, whence the assertion (i) . 
 
\medskip Let us now prove the assertion (ii); we know that the sequence $(\rho_\e)$ is bounded and therefore admits weak* cluster points. Given  such a cluster point, we can assume, without loss of generality,  that $\rho_\e \weak \rho$  in $\MM_+(\Ob)$.  
 Let us introduce for every $t>0$ the set 
 $$E_t:=\left\{ x\in \Ob \ :\ \liminf_{r\to 0}\, \frac{\rho(B(x,r))}{\omega_d r^d} > t  \right\}.$$
 Thanks to the upper semi-continuity of the map  $x\to \rho(B(x,r)$, we infer that $E_t$ is a Borel subset of $\Ob$.
We are going to prove that
\begin{equation}\label{claim1}
 \lim_{t\to +\infty}\,  \rho(E_t) =0  .
\end{equation}
  To that aim, we consider the family  of closed balls in $R^d$ defined by
$$ \FF_t  := \ \bigcup \left\{  \ov{B(x,r)} \ :\ x\in E_t,\ r< r_x ,\  \rho(\partial B(x,r))=0 \right\} ,$$
where $r_x>0$ is chosen so that $\rho(B(x,r))> t \omega_d r^d$ for every $r<r_x$.
Since $\FF_t$ determines a fine covering of the bounded Borel set $E_t$, we may invoke the Vitali-Besicovitch covering theorem (see \cite[Thm 2.19]{ambrosio}) which provides the existence a countable subfamily $(B_n)$ such that
\begin{equation}\label{Bn-cov}
 \rho(B_n)> t \, \LL^d(B_n) \  \forall n \quad,\quad  \rho(E_t\setminus \cup_n B_n)=0.
\end{equation}
Next we associate with the weak* convergent sequence $(\rho_\e)$,  two set functions defined on Borel subsets $A\subset \Ob$:
$$ \eta_\e(A):= F_\e   (\rho_\e\res A) \quad ,\quad \eta(A) :=  \liminf_{\e\to 0}  \eta_\e(A) .$$
We can readily check that $\eta_\e$ and $\eta$ are monotone with respect to the inclusion
while $\eta_\e(\Ob)= F_\e(\rho_\e) $ implies that $\eta(\Ob)\le \beta <+\infty$. Moreover, by the first inequality in \eqref{subp-eps}, $\eta_\e$ is super-additive on disjoint Borel subsets. Obviously this holds true also for the set function $\eta$. By applying this property to the sequence of disjoint balls $B_n$, we get
the upperbound:
\begin{equation}\label{sum-eta}  \sum_n \eta(B_n)  \ \le C\  .
\end{equation}
On the other hand, thanks to the coercivity inequality \eqref{ineq-coercive} that we apply with $\Ob=B_n$, we obtain:
\begin{align} \label{etaeps>} \eta_\e(B_n)= F_\e( \rho_\e \res B_n) \ \ge\  \ell_-(\d) \, \rho_\e(B_n) \left(  \dfrac{\rho_\e(B_n)}{\beta_\e(B_n)} -1 \right)_+ ,
\end{align}
where $\beta_\e (B_n) := \e^d \, m_\d \left( \dfrac {B_n}{\e} \right)$.
Since  $\rho(\partial B_n)=0$ by construction and thanks to \eqref{mdelta} and \eqref{Bn-cov},  we infer that
$$  \lim_{\e\to 0}  \dfrac{\rho_\e(B_n)}{\beta_\e(B_n)} \ =\ \dfrac{\rho(B_n)}{\d^{-d} \LL^d(\B_n)} \ \ge \ t\, \d^d  .$$
 Therefore, passing to the limit $\e\to 0$ in \eqref{etaeps>} , we deduce that: 
\begin{equation}\label{ineq-Bn}
 \eta(B_n) \ge \  \ell_- (\d)\ \rho(B_n)\ (t\, \d^d-1)_+ .
\end{equation}
 All in all, after collecting the second equality of \eqref{Bn-cov}, \eqref{sum-eta}  and  \eqref{ineq-Bn}, we are led to:
 $$ \rho(E_t) \ \le\  \sum_n \rho(B_n) \  \le\  \frac{C}{ \ell_-(\d) (t \d^d-1)_+ }\ .$$
Our claim \eqref{claim1} follows   by sending $t\to +\infty$.
The absolute continuity property $\rho\ll \LL^d$ stated in the assertion (ii)
is a  consequence of the \emph{Besicovitch  differentiation  theorem} \cite[theorem 2.22]{ambrosio}), which states that the singular part $\rho_s$ in the Lebesgue-Nikodym decomposition of $\rho$  with respect to the Lebesgue measure  
coincides with $\rho\res E_\infty$  being  $E_\infty=\cap_{t>0} E_t$. 
In our case  $\|\rho_s\|= \rho(E_\infty)=0$  due to \eqref{claim1}.

\end{proof} 

\subsection{Upper-bound of energies}
In the same way as in \cite{Lewin}, we will be using an upper bound  of $\EE_\ell(S)$ when $S$ is arranged on a $d$-dimensional 
 periodic Bravais lattice  $\GG$ 
\footnote{i.e. of the form  $\GG= F\, Z^d$ for some invertible matrix $F\in \R^{d\times d}$}. 
To such a lattice we associate the $\ell$- \emph{Epstein zeta function} defined 
 for every $r>0$ by:
\begin{equation}\label{def:Lll}
\Lambda_{\ell,\GG}(r) := \sum_{x\in \GG\setminus \{0\}}  \ \ell(r |x|).
\end{equation}
In the case where  the cartesian lattice $\GG=\Z^d$ is used, we will write simply $\Lambda_\ell(r)$. 
The finiteness of this function for large $r$, under the condition $(H3)$, turns out to be be crucial for
  deriving an  uniform upper bound for the scaled energy $F_\e$ given in \eqref{def:Feps}. 
  
\begin{lemma} \label{lattice}  Under  $(H1)-(H3)$, there exists  $C_\GG>0$ such that
\begin{equation}\label{zeta-bound}
\Lambda_{\ell,\GG}(r)\ \le\  \frac{C_\GG}{r^d} \, \left(\ell(r_0) \, r_0^d +  d \int_{r_0}^{+\infty} t^{d-1} \ell(t)\, dt\right) \quad \forall r \ge r_0\, \max \{1 ,a_\GG^{-1}\} \ , 
\end{equation}
 where $ a_\GG:=  \min\{|y|: y\in\GG\setminus \{0\}\}$.
\end{lemma}
 \begin{proof} \ To simplify, we chose the lattice  $\GG$ so that $a_\GG=1$. Up to substituting $\ell$ with $\ell_+$
 which satisfies $ \ell_+\ge \ell$ and  $\ell_+=\ell$ on $[r_0,+\infty)$, we may also assume that $\ell$ is non-increasing on $\R_+$. 
Accordingly, for any $s\in [0, \ell(0_+))$, the set of values $\{\ell >s\}$ forms a non-empty interval $[0,\ell^{-1}(s))$.
 The pseudo inverse $\ell^{-1}(s)$ is the supremum of all $t\ge 0$ such that $\ell(t)>s$, and it is a monotone non-increasing function on $[0,+\infty)$.
 Therefore, we have the following equivalence:
 $$\ell^{-1}(s)>t \iff \ell(t)>s .$$ 
 By applying the layer cake formula  to the counting measure on $\GG$, we get
 $$\Lambda_{\ell,\GG}(r)\ =\ \int_0^\infty  N_r(s)\, ds ,$$
 where  the integer function $N_r(s):= \sharp( \{x\in \GG\setminus \{0\} :  \ell(r|x|) >s\})$ satifies 
 $N_r(s)=0$ if $r\ge r_0$ and $s\ge \ell(r_0)$  (we assumed that $a_\GG=1$). 
 On the other hand, for any periodic Bravais lattice $\GG\subset \R^d$, there exists a constant $C_\GG>0$ such that
 $$   \sharp(B_r \cap \GG) \ \le\ C_\GG\,  r^d \quad, \quad \forall r>0 .$$
 This implies the inequality:
 $$ N_r(s) =\sharp( \{x\in \GG\setminus \{0\} : r |x| < \ell^{-1}(s)\})\ \le\  \frac{C_\GG}{r^d}\, (\ell^{-1}(s))^d .$$ 
Therefore, for every $r\ge r_0$, we are led to:
  $$\Lambda_{\ell,\GG}(r)\ \le\ \frac{C_\GG}{r^d}\, \int_0^{\ell(r_0)}  (\ell^{-1}(s))^d\, ds. $$
Then, after noticing that   $\{s\in [0,\ell(r_0)] \, :\, \ell^{-1}(s)>t\} = [0, \ell(r_0)\wedge\ell(t)] $
holds for any $t\ge 0$, we obtain the desired inequality by  applying once again the layer cake formula:
  \begin{align*} \int_0^{\ell(r_0)}   (\ell^{-1}(s))^d\, ds &= d \int_0^\infty  (\ell(r_0)\wedge\ell(t))\, t^{d-1}\, dt \\
  &=
  \ell(r_0) r_0^d + d \int_{r_0}^{+\infty} t^{d-1}\, \ell(t) dt. \end{align*}

 \end{proof} 
 
 \begin{remark}\label{jumps}
 In view of Lemma \ref{lattice}, the $\ell$- Epstein zeta function  $\Lambda_{\ell,\GG}(r) $ vanishes at infinity.
 However the behavior in $O(r^{-d})$  as $r\to\infty$ suggested by \eqref{zeta-bound} is not optimal 
as we can see in the case of a Riesz potential $\ell(r)=r^{-s}$ with $s>d$, where $\Lambda_{\ell,\GG}(r) =  C\, r^{-s}$.
On the other hand, it  is noteworthy that $\Lambda_{\ell,\GG} $ is not  continuous in general. A very simple example to see this is given by
the step function $\ell = \frac1{2} \, \one_{[0,1)}$ which satisfies $(H1)-(H3)$. For $d=1$ and the lattice $\GG=\Z$, we find that
$$\Lambda_\ell(r) =  \sharp \{ n\in \N: 0<n r <1\} = [r^{-1}]\ ,$$
 where $[\cdot]$ denotes the integer part.
 \end{remark}

\medskip
 Next, by applying Lemma \ref{lattice} in the case of the Cartesian lattice $\GG= \Z^d$, we 
derive a fundamental upper bound  for the short-range interaction energy. 
 
\begin{lemma}\label{upperb-lemma} Let $r>0$ and $S$ be a finite subset of the lattice $r \, \Z^d.$  
Then 
\begin{equation}\label{fundupper}
\EE_\ell(S)\ \le \  \sharp(S) \, \Lambda_\ell(r) \end{equation}
As a consequence, for every $a>0 $, there exists $S_\e\subset \Ob$ such that
$\rho_\e= \rho_{S_\e}^\e$ satisfies
\begin{equation}\label{epsupper}
\rho_\e \weak   a\, \LL^d\res \O\quad \text{and} \quad  \limsup_\e F_\e(\rho_\e) \le
 a \ \Lambda_\ell(a^{-\frac1{d}})  \, |\O| 
\end{equation}
where the right hand side upper bound is finite whenever $0\le a\le r_0^{-d}$.
  \end{lemma}

\begin{proof}  Let $S=\{ r\,x_i: 1\le i\le N\}$ where $N=\sharp(S)$ and $x_i\in \Z^d$.
 Noticing that, for every $i$,  the set $\{ x_i-x_j\, :\, j\not=i\}$  consists of $N-1$ distinct elements of 
 $\Z^d\setminus \{0\}$,
 we infer that
$$  \sum_{j\not=i}  \ell(r |x_i-x_j|)  \le   \sum_{z\in \Z^d\setminus \{0\}} \ell( r |z|)=\Lambda_\ell(r),$$
hence the desired inequality \eqref{fundupper}
by summing with respect to $i$.

\med
Taking now $\rho_\e= \rho_{S_\e}^\e$ where
$S_\e= \O\cap  (r_\e \Z)^d $ and $ r_\e= \e\, a^{-\frac{1}{d}}$,
we obtain a sequence such that $\rho_\e \weak a \, \LL^d\res \O$ as $\e\to 0$. Indeed, by the periodicity
of the Euclidean lattice, $\rho_\e$ converges to a uniform density on $\O$ while its total mass  $\|\rho_\e\|= \e^d \sharp(S_\e) 
 \sim  \e^d (r_\e^{-d} |\O|)$ converges  to $a \, |\O|$
as $\e\to 0$.
Eventually, by applying \eqref{fundupper}, 
we get $ F_\e(\rho_\e) = \e^d \EE_\ell(\frac{S_\e}{\e}) \le \e^d\, N_\e\, \Lambda_\ell(a^{-1/d}) ,$
whence:
$$  \limsup_\e  F_\e(\rho_\e) \le   a \ \Lambda_\ell(a^{-\frac1{d}})  \, |\O|.$$
 The finiteness of  $\Lambda_\ell(a^{-\frac1{d}}) $ for  $a\in [0,  r_0^{-d}]$  follows from Lemma
 \ref{lattice}.

\end{proof}

\begin{remark}\label{opti-lattice}\ The upper bound \eqref{fundupper} and \eqref{epsupper} obtained by choosing $\GG=\Z^d$ as the reference lattice are in general not optimal since the Epstein-zeta function of another lattice could provide  better ones. Obviously the inequality \eqref{fundupper} holds true after replacing  $\Lambda_\ell$ with the Epstein function $\Lambda_{\ell, \GG}$ of any Bravais lattice $\GG= F \, \Z^d$, while for the valisity of  \eqref{fundupper}, we need to add  the normalization condition $|{\rm det} (F)|=1 $
(thereby fixing the volume of the so called \emph{fundamental domain} of $\GG$). 
The existence and the determination of an optimal lattice $\GG$ for the following minimization problem:
 $$
\inf\{\Lambda_{\ell,\GG}(r): \GG= F\, Z^d,\ |{\rm det} (F)|=1 \}
 $$
touches on a very hard and famous problem related to crystallisation conjectures (see \cite{Lewin, Lewin-Blanc}). 
Note that, for a general cost $\ell$, the answer to this problem will depend of the value $r$ (thus of the local density $a= \dfrac{d\rho}{dx}$ of the limiting measure $\rho$).
\end{remark}

\section{The $\Gamma$-convergence result.}  

\medskip

\subsection{A quick overlook}\ The notion of $\Gamma$-convergence is popular in the community of calculus of variations and very much used in the analysis of sharp-interface models, dimension reduction for  problems in mechanics, optimal design and homogenization. As pointed out in the introduction, this tool is also perfectly suited to justify a mean-field approach for large particle systems subject to a minimum energy criterion (see \cite{serfaty2018systems}). For the convenience of the reader, let us give here some basic definitions  and main properties. For futher details, we  refer to the monographs \cite{attouch1984, dal2012, braides2002gamma}.
 
\med
 Let $(E,\tau)$ be a metrizable topological space and consider a sequence of functionals $F_n: E \to (-\infty,+\infty]$. Then the lower
 $\Gamma$-limit $F_-$ and the upper $\Gamma$-limit $F_+$  of $F_n$ are defined by:
 $$    F_-(u) := \inf_{u_n\to u} \liminf_{n\to\infty}  F_n(u_n)\quad,\quad 
  F_+(u) := \inf_{u_n\to u} \limsup_{n\to\infty}  F_n(u_n) .$$
  Both are $\tau$-lower semicontinuous  (see \cite{attouch1984}),  whereas  in general it holds $F_-\le F_+$. 
  In practice it is useful to check that these functionals are proper i.e. that they range into
  $\R\cup \{+\infty\}$  being not identically $+\infty$. If $F_n$ admits a  lower bound independent of $n$, this amounts to checking that
  the existence of $u_0\in E$ such that
  \begin{equation}\label{proper}
 F_+(u_0)= \inf_{u_n\to u_0 } \limsup_{n\to\infty}  F_n(u_n)<+\infty .
\end{equation}

  We say that $F_n$ $\Gamma$-converges to $F$  (denoted $F_n\Gconv F$) if $F=F_-=F_+$ or equivalently 
if the two following conditions are fulfilled: 
\begin{itemize}
\item[a)] \ \emph{(lowerbound)}  \  For any sequence $u_n$ converging to $u$, we have the inequality
$$\liminf_{n\to \infty} F_n(u_n) \ge F(u) ;$$
\item[b)] \ \emph{(recovering sequence)}  \ For every $u\in X$ such that $F(u)<+\infty$, there exists $(u_n)$
such that  $$ u_n\to u\quad \text{and} \quad F_n(u_n)\to F(u).$$
\end{itemize}
This is the case in particular if $F_n=F$ does not depend of $n$; then $F_n\Gconv \cl(F)$ where $\cl(F)$ denotes the $\tau$-lower semicontinuous envelope of $F$.
Among all properties of $\Gamma$-convergence, we give some which will be used in this paper.
\begin{prop}\label{essential}  \ Let $F_n: E \to (-\infty,+\infty]$ and assume \eqref{proper}.
Then:
\quad 
\begin{itemize}

\item  [(i)]\  $F_n \Gconv F  \iff  \cl(F_n) \Gconv F$;

\item  [(ii)]\   (\emph{Kuratowski compactness Theorem}) \ If $(E,\tau)$ is a second countable topological space (for instance a separable metric space), then any sequence $(F_n)$ admits a $\Gamma$-convergent subsequence;

\item  [(iii)]\ (\emph{convergence of infima}) Suppose that $F_n \Gconv F$ and that the following equi-coercivity property holds:
 $$\sup_n F_n(u_n)<+\infty \implies \{u_n\} \text{is $\tau$-relatively compact} .$$
 Then $\lim_{n\to \infty} \inf_X F_n =  \min_X F$ and the minimum set for $F$ coincides with the cluster points of all sequences $(u_n)$ such that $F_n(u_n) - \inf F_n \to 0$ ;
\item  [(iv)]\  (\emph{stability}) \ $F_n \Gconv F  \implies  F_n+G \Gconv F+G$  for every continuous perturbation function $G:E \to \R$.  
\end{itemize}
\end{prop}

\begin{remark} \label{G-extend} The continuity requirement for  $G$ in the  assertion (iv) is often too restrictive.
Actually the same conclusion holds under the following milder condition:
\begin{equation} \label{stability}
\begin{cases} \ds  \inf_K G > -\infty  & \text{for any compact $K\subset E$}\\  
 G(u_n) \to G(u) & \text{ whenever $u_n \to u$ and $F(u)<+\infty$} 
 \end{cases}
\end{equation}
For the convenience of the reader, a brief proof of the sufficiency of this condition is given below.
\begin{proof} To check condition a), we consider a sequence $(u_n)$ such that $u_n\to u$. Without loss of generality, assume that
$F_n(u_n) + G(u_n) \le C$ for a suitable constant $C$. Since $G(u_n)$ is lower bounded (take $K$ to be $\{u_n, n\in\N\} \cup \{u\}$),
we infer that $F_n(u_n) \le C'$ for another constant $C'$. By the $\Gamma$-convergence $F_n\Gconv F$, it follows that
$$F(u) \le  \liminf_n F_n(u_n) <+\infty . $$
Therefore $G(u_n) \to G(u)$ and $\liminf_n (F_n+G)(u_n) \ge (F+G)(u)$.

\medskip
For checking condition b), we may restrict to elements $u\in E$ such that $(F+G) (u)<+\infty$. Then $F(u)<+\infty$  and any recovering sequence $u_n\to u$  such that $F_n(u_n)\to F(u)$ will satisfy $G(u_n)\to G(u)$. 
\end{proof}

%If the sequence $(F_n)$ is equi-coercive in the dual of
%a separable Banach space $X$, then the lower and upper $\Gamma$-limits are unchanged if we restrict $F_n$ to a closed ball of $X^*$  
%which is weakly* compact and metrizable; therefore Kuratowski compactness theorem still applies to this case when $\tau$ is the weak* topology. 
%%In our case $E$ will be a closed ball of $\MM(\Ob)$ embedded with the weak* topology. 
\end{remark}
\subsection{ The main result}  
In our  context, the $\Gamma$-convergence issue applies to the sequence $(F_\e)$ defined in  \eqref{def:Feps}  and to the ambiant topological space  $\MM_+(\Ob)$ embedded with the weak* topology (tight convergence). Thanks to  the equi-coercivity property established in Proposition \ref{compactness-Feps}, there is no loss of generality in working
in a fixed closed ball of $\MM_+(\Ob)$  which is metrizable and compact. Therefore all the properties mentioned in the former subsection are applicable (after substituting  the index $n\to \infty$ with the continuous parameter $\e\to 0$).
 
 \begin{thm}\label{main-thm} \  Let $\ell$ satisfy the standing assumptions $(H1)-(H3)$ and 
 let $F_\e: \MM_+(\Ob) \mapsto [0,+\infty]$ be given by \eqref{def:Feps}. Then 
 $  F_\e   \ \Gconv F $  (for the weak* topology) where
$$ F(\rho) :=  \begin{cases} \int_\O  f_\ell (u)\, dx & \text{if $\rho= u\, \LL^d \res \O$}\\
 +\infty & \text{otherwise} \end{cases} $$
and  $f_\ell: \R_+ \to [0,+\infty]$ is convex, l.s.c. and satisfies
\begin{equation}\label{quadra}
f_\ell(0)= f'_\ell(0_+) =0\quad, \quad 
  \liminf_{t\to +\infty} \frac{f_\ell(t)}{t^2} >0. 
\end{equation}

 \end{thm}

The proof of Theorem \ref{main-thm} is postponed to Section 5 while the property \eqref{quadra} is established in the next Sub-section.
 Let us now consider the Fenchel conjugate of $f_\ell$
(implicitly extended by $+\infty$ on $(-\infty,0)$) given by
\begin{equation}\label{fell*}
f_\ell^*(\la) \ :=\  \sup \left\{  \la \, t - f_\ell(t) \ :\ t\in \R_+\right\}
\end{equation}
From \eqref{quadra}, one can check that the supremum in \eqref{fell*} is actually a maximum 
(which is attained at $t=0$ if $\la \le 0$).
Therefore, $f_\ell^*$ is convex,  continuous and vanishes  on $(-\infty, 0]$.
As a consequence it admits left and right derivatives $ (f_\ell^*)'(\la_-) \le (f_\ell^*)'(\la_+)$ for any $\la$
so that the subdifferential $\partial f_\ell^*(\la)=[(f_\ell^*)'(\la_-),(f_\ell)^*)'(\la_+)]$ is non empty. 
%%%%
\begin{corollary}\label{mean-field} Let $U\in \CC(\Ob)$ be an external potential. Then, for every $\e>0$, there exists a finite set $S_\e\subset \Ob$
minimizing
$$\I_\ell^{(\e)}(U) :=  \inf_{S\subset \Ob} \Big\{ \sum_{(x,y)\in S^2\setminus \Delta} \ell\left(\frac{|x-y|}{\e}\right) +  \sum_{x\in S} U(x) \Big\}.$$
Moreover $ \sup_\e \e^d \, \sharp(S_\e) <+\infty $  and 
$  \lim_{\e\to 0} \, \e^d\, \I_\ell^{(\e)}(U) \, =\, \I_\ell(U)$, where
\begin{equation}\label{mean-pb}
\I_\ell(U) := \min_{u\in L^1(\O)} \left\{ \int_\O (f_\ell(u) + u\,U)\, dx  \right\}  \, =\,  -\int_\O f_\ell^*(-U) \, dx.
\end{equation}
 Furthermore any  weak* cluster point of $ \rho_{S_\e}^\e$ 
 belongs to the minimum set of \eqref{mean-pb} given by 
\begin{equation}\label{mean-set} \SS_\ell:= \left\{ u \, \LL^d\res\O \ :\  u(x)\in \partial f_\ell^* (-U(x)) \ \text{a.e.} \, x\in\O \right\}. \end{equation}

\end{corollary}
\begin{remark} [\emph{Non-uniqueness}] \label{nonunique} \
In general $f_\ell$ is not strictly convex and $\partial f_\ell^*$ can be multi-valued (see for instance the hard spheres case depicted in Subsection \ref{examples} or the example of the step function $\ell$ given in \eqref{step-ell} where $f_\ell$ is piecewise affine). 
Note that, since $f_\ell^*$ vanishes on $\R_-$, it holds $\I_\ell(U)=0$  for every non-negative potential $U$.

\end{remark}

\begin{remark} \label{variant}
   A natural variant of $\I_\ell^{(\e)}(U)$ consists in prescribing the total number  $N_\e$ of particles 
to satisfy  $N_\e \sim \kappa \, \e^{-d}$ for some given intensity factor $\kappa\in (0,+\infty)$. 
 Accordingly, Corollary \ref{mean-field} can be restated by adding a total mass constraint in the 
   limit problem, that is with  $I_\ell(U)$ in \eqref{mean-pb} replaced by
$$ \I_{\ell,\kappa} (U) := \inf \left\{ \int_\O (f_\ell(u) + u\,U)\, dx \, :\, u\in L^1(\O;\R_+) \ ,\ \int_\O u\, dx = \kappa\right\}.$$
The associated minimum $\SS_{\ell,\kappa}$ can be determined by selecting  a suitable Lagrange multiplier depending implicitly on $\kappa$; as a consequence, an explicit form for $\SS_{\ell,\kappa}$
of the kind \eqref{mean-set} is not available. 
\end{remark}

\begin{remark}[\emph{Clustering}] \label{relax}  \  For  $\e>0$ fixed, the lower semicontinuity property of $F_\e$ requires that $\ell(0+)=+\infty$. Otherwise,  if $\ell(0_+)<+\infty$,
a sequence of subsets $S_n\subset\Ob$ such that $\sup_n F_\e(\rho_{S_n}^\e)<+\infty$  (thus retaining a finite number of points) can collapse into several clusters while retaining finite energy. In this case, the relaxed functional $\cl(F_\e)$ can be obtained directly by extending its domain to $\e$-empirical measures associated with  multisets (instead of sets) 
and by extending the definition \eqref{eps-muS} accordingly  taking into account the multiplicity of each cluster of particles.
Note however that considering   $\cl(F_\e)$ instead of $F_\e$ will not change the mean-field energy $F$ given in Theorem \ref{main-thm}
in virtue of the assertion i) of Proposition \ref{essential}.

 \end{remark}

\subsection{Characterization and properties of $f_\ell$}\  The convex integrand $f_\ell$ will be characterized indirectly through its Fenchel conjugate. For every $\la\in\R$ and any Borel subset $B\subset \R^d$, we define:
\begin{equation}\label{def:Gamma-ell}
\Gamma_\ell(\la,B) \ :=\ \sup \left\{ \la\, \sharp(S) - \EE_\ell(S) \ :\ S \ \text{finite $\subset B$} \right\}.
\end{equation}
The key properties of this bivariate function are summarized in the two following lemmas. 
Let $\f:\R \to \R$ be the convex, continuous function defined by:
\begin{equation}\label{def:fi}
 \f(t)=\begin{cases} \frac{(1+t)^2}{4} & \text{if $t\ge1$}\\ t & \text{if $0\le t< 1$}
 \\ 0 & \text{if $t<0$} \end{cases}.
\end{equation}
A straightforward computation shows that its Fenchel conjugate is given by:
$$ \f^*(\zeta) = \zeta (\zeta-1)_+  \quad \text{if $\zeta\ge 0$} \quad,\quad \f^*(\zeta)=+\infty\quad \text{if $\zeta< 0$}.$$

\begin{lemma}\label{fl1} Assume that $\LL^d(B)<+\infty$. Then the map
$\la \to \Gamma_\ell(\la,B)$ is convex, continuous, vanishes on $\R_-$ and satisfies:
\begin{equation}\label{upperGammell}
 0 \, \le\,  \Gamma_\ell(\la,B)\ \le\   m_\d(B) \ \ell_-(\d)\, \f \left(\frac{\la}{\ell_-(\d)}\right) , \end{equation}
for every $\la\in \R$, being $\d>0$ such that $\ell_-(\d)\in (0,+\infty]$ and $\f$ being defined by \eqref{def:fi}.

\end{lemma}
Notice that the right hand side of \eqref{upperGammell} is convex as a function $\la$.
On the other hand, due to the linear behavior of $\f$ on $[0,1]$,  we can easily infer that  $\Gamma_\ell(\la,B)\le \la\,  m_\d(B)\, $
whenever $\ell_-(\d)=+\infty$.
\begin{proof} As a supremum of affine functions, $\Gamma_\ell(\la,B)$ is convex and l.s.c. with respect to $\la$;
it is non-negative (follows by taking $S$ the empty set) and vanishes if $\la\le 0$. 
The continuity property follows classically from  \eqref{upperGammell} which provides the finiteness  of  $\Gamma_\ell(\la,B)$
since $m_\d(B)<+\infty$ once $\LL^d(B)<+\infty$. 
It remains to prove the upper bound in \eqref{upperGammell} which clearly follows from \eqref{fundlower}.
Indeed, in term of  $\zeta= \frac{\sharp(S)}{m_\d(B)}$, we have for every finite $S\subset B$:
$$ \la\, \sharp(S) - \EE_\ell(S)\ \le\  m_\d(B)\,  \left( \la \, \zeta - \ell_-(\d) \, \zeta\, (\zeta-1)_+\right) .$$
Taking the supremum of the right hand member with respect to $\zeta\ge 0$ and noticing that $\zeta\, (\zeta-1)_+=\f^*(\zeta)$, we derive \eqref{upperGammell}after some easy manipulations. 

\end{proof}

\begin{lemma}\label{fl2} For every $\la\ge 0$, the set function 
 $B\mapsto \Gamma_\ell(\la,B)$ is subadditive on disjoints Borel subsets and translation invariant.
\end{lemma}
\begin{proof} 
Let $B_1,B_2$ be such that $B_1\cap B_2=\emptyset$ and let $S$ be a finite subset of $B_1\cup B_2$.
Then, setting $S_i= S\cap B_i$, we get a partition $S=S_1\cup S_2$ and by appling the super-additivity part of \eqref{subp}, we infer that:
$$ \la \, \sharp(S) - \EE_\ell(S) \le  \sum_{i=1}^2  (\la \, \sharp(S_i) - \EE_\ell(S_i)) \le \Gamma_\ell(\la,B_1)+\Gamma_\ell(\la,B_2) ,$$
hence the desired sub-additivity property by taking the supremum with respect to $S\subset B_1\cup B_2$.
The invariance by translation is trivial.

\end{proof}

In virtue of Lemma \ref{fl2},  we may now  apply to $\Gamma(\la,\cdot)$ a classical result by Krengel \cite{Krengel},  which ensures the existence, for every $\la$, of a  limit for the ratio $\frac{\Gamma(\la, Q_k)}{k^d}$ as $k\to+\infty$ (\emph{thermodynamical limit}).
 Let us define the function 
 \begin{equation}\label{def:gell}
g_\ell(\la):=  \inf_{k\in \N} \frac{\Gamma(\la, Q_k)}{k^d} .
\end{equation}
Then we have:
\begin{equation}\label{gellim}
g_\ell(\la)= \, \lim_{k\to +\infty} \frac{\Gamma(\la, Q_k)}{k^d}\,  =\,  \lim_{\e\to 0} \e^d \, \Gamma(\la, Q_{1/\e})\,.
\end{equation}
For a proof of \eqref{gellim}, we refer for instance to \cite{licht2002global}.  Since the effective profile $f_\ell$ in Theorem \ref{main-thm}  will be identified through the relation  $f_\ell^*=g_\ell$
 (see the last step of the proof in Section 5), we now establish some useful bounds for $g_\ell$.
 Recalling the definition of $\Lambda_\ell$ in \eqref{def:Lll}, we introduce the function defined on $\R$ by:
 $$ H_\ell(t):=\begin{cases}  t\, \Lambda_\ell ( t^{- \frac{1}{d}}) & \text{if $t>0$ \,,} \\ +\infty & \text{if $t\le0$} \,.\end{cases} $$
From $(H_1)(H_2)$, it is straightforward that this function  $H_\ell$ is  l.s.c. and finite  on $(0, r_0^{-d}]$ while
$ H_\ell(0_+)= H_\ell'(0_+)=0$.
Note that $H_\ell$ is not convex in general, even it could be discontinuous, as 
happens for $\ell$ which is the step function
$\ell(r)= \frac1{2} \, \one_{[0,1)}$. Indeed, in this case and for $d=1$, we get $H_\ell (t) = t \, [t]$ (see Remark \ref{jumps}).
However $H_\ell$ is convex continuous on $\R_+$  in many classical cases including
that  of hyper singular Riesz potentials $\ell(r)=r^{-s}$ where  $ H_\ell(t)= C\, t^{1+ s/d}$.

 \begin{prop}\label{bounds-gell} The fonction $g_\ell(\la)$ is convex, continuous, non-negative, vanishes on  $\R_-$ and, for $\la>0$, satisfies the inequalities: 
\begin{equation}\label{lowupgell}
 H_\ell^*(\la)\ \le\ g_\ell(\la)\ \le\  \d^{-d} \ \ell_-(\d)\ \f \left(\frac{\la}{\ell_-(\d)}\right) , \end{equation}
 being $\f$  given by \eqref{def:fi} and $\d>0$ choosen such that $\ell_-(\d)>0$.  \footnote{The second inequality becomes
 $ g_\ell(\la) \le\  \d^{-d}\, \la$ in the case where $l_-(\d)=+\infty$. }
\end{prop}

\begin{proof} In view of \eqref{gellim} and of the convexity of $\Gamma(\cdot, Q_k)$, the function $g_\la$ is convex, non-negative and vanishes for $\la\le 0$ as a pointwise limit as $k\to\infty$  of the sequence of functions $k^{-d} \, \Gamma(\cdot,Q_k)$.  

To prove the right hand inequality in \eqref{lowupgell}, it is enough to apply \eqref{upperGammell} with $B=Q_k$ and, after dividing by $k^d$, pass to the limit taking into account that, in virtue of \eqref{mdelta},  $m_\d(Q_k)\sim (\frac{k}{\d})^d$ as $k\to\infty$.  Since the majorant is a convex, continuous function of $\la$, we infer that $g_\ell$ is continuous on $\R$ as well.

Eventually, let us apply, for every $t>0$ anf $k\in \N_*$, the fundamental upper bound \eqref{fundupper} to the finite subset $S_{t,k}:= t^{-\frac1{d}} Z^d \cap Q_k .$ Then, recalling \eqref{def:Gamma-ell}, we have:
$$  \Gamma_\ell(\la, Q_k)\ \ge \ \la \, \sharp(S_{t,k}) - \EE_\ell (S_{t,k}) \ \ge\ \sharp(S_{t,k}) \left(\la- \Lambda_\ell ( t^{- \frac{1}{d}})\right).$$
Since $\sharp(S_{t,k}) \sim t\, k^d$ as $k\to \infty$ and in virtue of \eqref{gellim}, we are led to: 
$$g_\ell(\la) = \lim_{k\to \infty} \frac {\Gamma_\ell(\la, Q_k)}{k^d}\ \ge \  t (\la  -\Lambda_\ell ( t^{- \frac{1}{d}}))\ =\ 
\la t - H_\ell(t) .$$
The left hand side inequality in \eqref{lowupgell}  follows by taking the supremum with respect to $t$.
\end{proof}

 \begin{corollary}\label{bounds-fell} The integrand $f_\ell:= g_\ell^{*}$ is convex, l.s.c. and satisfies $f_\ell(0)=0$ while $f_\ell(t)=+\infty$ for $t< 0$. Moreover, for every $t>0$ and $\d>0$, we have the inequalities: 
\begin{equation}\label{lowupfell}
 \ell_-(\d) \ t\, (t\, \d^{d} -1)_+\,  \le\, f_\ell(t) \ \le\  H_\ell^{**}(t) . \end{equation}
Accordingly, $f_\ell$ is finite on $ [0,r_0^{-d}]$, monotone non-decreasing on $[0,+\infty)$ and  satisfies: 
\begin{equation}\label{superquadra} f_\ell'(0_+)=0 \quad,\quad \liminf_{t\to +\infty}\ \frac{f_\ell(t)}{t^2}\, \ge\,  \sup_{\d>0}\, \ell_-(\d) \d^d > 0\end{equation}
\end{corollary}
\begin{proof} Passing carefully to Fenchel conjugates in inequalities \eqref{lowupgell}, we are led to \eqref{lowupfell}
from which the other statements follow directly. In particular, as $f_\ell'(0_+)=0$, we infer  that the convex function
$f_\ell$ is monotone non-decreasing on $\R_+$.   
\end{proof}

\begin{remark}  [\emph{growth conditions}]  \ The inequality in \eqref{superquadra} confirms that $f_\ell$ grows at least quadratically at infinity, as  announced in the
introduction (see Theorem \ref{main-thm}). More specifically, we can highlight two subcases for a cost $\ell$ satisfying1 $(H1)-(H_3)$.
\setlength{\leftmargini}{10pt}
\begin{itemize}\item [a)]  $\ds k_\ell:= \sup_{\d>0}\, \ell_-(\d) \d^d =+\infty.$ \ Then
 $ \liminf_{t\to +\infty}\ \frac{f_\ell(t)}{t^2} =+\infty$ and $f_\ell$ has a super quadratic growth. Note that this conclusion is consistent with the case $\ell(r)=r^{-s}$ for  $s>d $ (see the next subsection).

\medskip
\item [b)]  $\int_0^{\infty} \ell_+(t) t^{d-1}\, dt<+\infty.$ \  In this case $k_\ell<+\infty$ and, thanks to \eqref{lowupfell} and to the estimate  
given in Lemma \ref{lattice} (that we can apply to $\ell_+$ with $r_0= 0$), we obtain the lower and upper bounds:
$$  0< k_\ell \le  \liminf_{t\to +\infty}\ \frac{f_\ell(t)}{t^2} \le  \limsup_{t\to +\infty}\ \frac{f_\ell(t)}{t^2} \le 
C \int_0^\infty \ell_+(t) t^{d-1}\, dt . $$
It follows that, under the integrability condition $\int_{\R^d} \ell_+(|x|) \, dx<+\infty$,  $f_\ell$ enjoys a quadratic growth from above and from below.

\end{itemize}

\end{remark}

\medskip

\subsection{Examples } \label{examples}

\subsubsection{The hard spheres model}\label{hard} \ The hard spheres potential is given by  
$$\ell(r)= \begin{cases} +\infty & \text{ if $r<1$} \\
0 & \text{ if $r<1$} \end{cases}$$
The computation of $g_\ell$ through \eqref{def:gell} and \eqref{gellim} leads to a linear function on $\R_+$ namely $g_\ell(\la) = \gamma_d\, \la$,  where  $\gamma_d$  denotes
the densest spheres packing volume fraction in $\R^d$.  This famous universal constant can be defined as
 \begin{equation} \label{def:packing}
	\gamma_d := \inf_{k\in \N^*} \frac{S(Q_k)}{k^d} = \lim_{k\to \infty} \frac{S(Q_k)}{k^d},
\end{equation}
 where, for any Borel set $A\subset\R^d$, $S(A)$ denotes the maximal number of points in $A$ with mutual distance larger or equal to $1$.
 The mean-field energy density $f_\ell$ given by Theorem \ref{main-thm} is therefore the indicator function of the interval $[0,\gamma_d]$
 $$  f_\ell(t) = 0\quad\text{if $t\le \gamma_d$} \quad, \quad f_\ell(t) = +\infty   \quad \text{otherwise}\ .$$
 Furthermore, for every continuous external potential $U\in \CC(\Ob)$, we recover from Corollary \ref{mean-field} the convergence:
 $$  \min_{S\in \FF_\e(\Ob)} \left\{\e^d\,   \sum_{z\in S}  U(z)  \right\} 
 \to \gamma_d\, \int_{\Ob} U(z)\, dz \ ,$$ 
 where $\FF_\e(\Ob)$ is the family of finite subsets $S\subset  \Ob$ satisfying  $|x-y| \ge \e $ for all $(x,y)\in S^2\setminus \Delta$.
 
 \begin{remark}\label{filling} A variant of the previous result was obtained recently in \cite{BindBou2022} in the case where the total number of particles $N_\e$ is prescribed to satisfy $N_\e \, \e^d \to \kappa $ as $\e\to 0$ where $\kappa>0$ is a given real parameter. With our notations this condition
 amounts to restrict the $\Gamma$-limit $F$ to  measures $\rho$ such that $\int_\O udx =\kappa$.
 Since  the domain of $F$ consists of density measures $\rho= u \, \LL^d\res\O$ such that $u\le \gamma_d$ a.e., the latter integral condition requires that $\kappa \le \gamma_d \, |\O|$
 hence a congestion ratio $\theta:= \frac{\kappa}{\gamma_d |\O|}$ not larger than $1$. In this case and if, following the classical empirical measure representation,  $u$ is normalized to be a probability density by setting ${\tilde u}:= \frac{u}{\int_\O udx}= \frac{u}{\kappa}$, we recover a mean-field energy vanishing for ${\tilde u}\le \frac{\gamma_d}{\kappa}$  and infinite otherwise, exactly as stated in \cite[Thm 6.1]{BindBou2022}.
 Note that the duality technique used there could only handle cost funtions $\ell$ taking values in the discrete set $\{0, +\infty\}$.  
 \end{remark}

\subsubsection{The case of Riesz potentials} Short range potential of Riesz type corresponds to fixing $s>d$ and taking
\begin{equation*}
l(r)=r^{-s} \text{ on } \mathbb{R}^d_+\,.
\end{equation*}
In this case, it's easy to establish from the homogeneity of the cost  $\ell$ that the bivariate function $\Gamma_\ell$ defined in \eqref{def:Gamma-ell} satisfies, for every $t\ge 0$, the following scaling law :
\begin{equation}\label{scaling-Gamma}
\Gamma_\ell(t \lambda,B)= t \Gamma_\ell(\lambda, t^{1/s} B)\,.
\end{equation}
It follows from \eqref{scaling-Gamma} that:
\begin{align*}
 \dfrac{\Gamma_\ell(t \lambda, Q_k)}{k^d} =
 \dfrac{t \Gamma_\ell( \lambda, t^{1/s}Q_k)}{k^d}=
 t^{1+d/s} \, \dfrac{ \Gamma_\ell( \lambda, Q_{t^{1/s}k})}{(t^{1/s}k)^d} .
\end{align*}
%\begin{align*}
%g_\ell(t \lambda)=\lim_{k\to \infty} \dfrac{\Gamma_\ell(t \lambda, Q_k)}{k^d} =
%\inf_k \dfrac{t \Gamma_\ell( \lambda, t^{1/s}Q_k)}{k^d}=
%\inf_k \dfrac{t \, t^{d/s}\Gamma_\ell( \lambda, Q_{t^{1/s}k})}{(t^{1/s}k)^d}
%=t^{1+d/s}\, g_\ell(\lambda) .
%\end{align*}
Sending $k\to\infty$ and applying \eqref{gellim} two times,  we get:
$$ g_\ell(t \lambda)=\lim_{k\to \infty} \dfrac{\Gamma_\ell(t \lambda,  Q_k)}{k^d}=
\lim_{k\to \infty} t^{1+d/s} \dfrac{\Gamma_\ell( \lambda, Q_{t^{1/s}k})}{(t^{1/s}k)^d}=t^{1+d/s}\, g_\ell(\lambda). $$
In virtue of the equality $f_\ell= g_\ell^*$, we deduce that:
\begin{equation}
f_\ell(t)=  C(s,d)\, t^{1+s/d}
\end{equation}
where $C(s,d)=f_\ell(1)$ is a universal constant. We thus recover 
the $\Gamma$-convergence result proved in  \cite{hardin2021, HardSerfLebl}.

\begin{remark}\label{lim=infty}\ If we chose $\ell(r)=r^{-s}$ where $s<d$, then condition $(H3)$ is violated, and the scaling defined in equation \eqref{def:Feps} that we used to define $F_\e$ will result in an infinite $\Gamma$-limit. 
 This means that $F(\rho)$ will be equal to $+\infty$ whenever $\rho$ is not equal to zero, and $F(0)$ will be equal to zero. 
We can observe this when we consider a system of $N_\e$ particles in $S_\e\subset \Ob$ such that $\rho_\e:= \rho_{S_\e}^\e$ converges weakly to $\rho$, and $\sup_\e F_\e(\rho_\e) < +\infty$.
Assuming $\rho\neq0$, then we have $N_\e \sim \|\rho\|\, \e^{-d}$ as $\e \to 0$. Moreover, due to the power law property of $\ell$, we can write:
\begin{equation}\label{equival}
F_\e(\rho_\e) = \e^{s+d}\, \EE_\ell(S_\e) \sim \frac{\|\rho\|^2}{\e^{d-s}} 
\, \frac{\EE_\ell(S_\e)}{N_\e^2}. \end{equation}
As $\ell$ satisfies \eqref{L1loc}, the convergence result of the long range case holds (with $h_N=N^{-2}$, see \cite{BindBou2022}, \cite{Serfaty2015}). Therefore, based on \eqref{def:Dell}, and given that the standard empirical measure linked to $S_\e$ converges to $\hat \rho=\frac{\rho}{\|\rho\|}\in \PP(\Ob)$, it follows that 
$$ 
\liminf_{\e\to 0} \frac{\EE_\ell(S_\e)}{N_\e^2} \ge D_\ell(\hat \rho) >0
.$$ 
This contradicts \eqref{equival} since $\sup_\e F_\e(\rho_\e)<\infty$. Therefore $\rho=0$.\end{remark}

\subsubsection{ The case of finite costs}\  Many examples of finite costs can be considered as, for instance, $\ell$ being a step function with compact support. Owing to Corollary  \ref{bounds-fell}, the effective convex integrand $f_\ell$ has a quadratic growth on $\R_+$.
The simplest one is  the penalized version of the hard spheres potential defined by:
\begin{equation}\label{step-ell}
\ell(r)= \begin{cases} \frac{M}{2} & \text{ if $r<1$} \\
0 & \text{ if $r<1$} \end{cases}\qquad\text{ ($M$ positive parameter) }.
\end{equation}

Applying the lower bound \eqref{lowupfell} with $\delta=1$, we deduce that $f_\ell  \ge h$
where
$$h(t):= \frac{M}{2} t (t-1)_+ .$$
 In turn this lower bound is optimal for integer values of $t$  
since, as proved below, $f_\ell$ coincides on $\R_+$ with the piecewise affine interpolation of $h$ given by:
\begin{equation}\label{fMexpli}
f_\ell(t) =h(k)  + (t-k) (h(k+1)-h(k))\quad \forall t\in [k,k+1]\ , k\in \N . 
\end{equation}
 \begin{proof} Owing to \eqref{gellim}, the Fenchel conjugate of $f_\ell$ is given by
$$ g_\ell(\lambda)= \lim_{K\to +\infty} \frac1{K} \ \sup_{S\subset [0,K]}
\left\{\lambda \ \sharp(S) - \frac{M}{2} \, \sharp (\{(x,y)\in S^2\setminus\Delta_1\}) \right\},$$
where $\Delta_1:= \{(x,y) \in\R^d\times \R^d : |x-y|\ge 1\}$.
Let $S\subset [0,K]$ be an optimal set which we split in $K$ disjoint pieces namely
 $S = \bigcup_{i=1}^K  S_i$ where $S_i = S\cap [i,i+1) .$
 Let us denote $n_i$ the number of points in $S_i$.
 By pushing them to the center of the interval $[i,i+1]$, we see that the number of pairs in $S^2\setminus\Delta_1$ decreases to 
 $ \sum_{i=1}^K n_i(n_i-1)$. It follows that:  
%\begin{align*}
%  g_\ell(\lambda) &= \lim_{K\to\infty}  \sup_{n_i\in \N} \frac1{K}  \left\{ \la\ \sum_{i=1}^K n_i - \frac1{2} \sum_{i=1}^K n_i(n_i-1) \right\} \\
%&= \sup_{n\in \N}  \left\{\la\, n -  \frac1{2} n(n-1)\right\} \, =\, \sup_{n\in \N}  \left\{(\la+ \frac1{2}) n - \frac1{2} n^2\right\} . 
%\end{align*}
\begin{align*}
  g_\ell(\lambda) &= \lim_{K\to\infty}  \sup_{n_i\in \N} \frac1{K}  \left\{ \la\ \sum_{i=1}^K n_i - \frac{M}{2} \sum_{i=1}^K n_i(n_i-1) \right\} \\
&= \sup_{n\in \N}  \left\{\la\, n -  \frac{M}{2} n(n-1)\right\} \, =\, (h + \chi_{\N})^*(\la) , 
\end{align*}
where $\chi_{\N}$ denotes the indicator function of the integers. Therefore $f_\ell= (g_\ell)^*$ is nothing else but the convexification of 
$h + \chi_{\N}$ given by the interpolation formula \eqref{fMexpli}.
\end{proof}
% The last supremum taken on all $\R_+$ coincides with $h^*(\la) = \frac1{2} (\la+ \frac1{2})^2$. Restricting to the integers, we get an integrand $g_\ell(\la)$ \todo{a revoir} which is  affine on $(0,1)$ with slope $g_\ell'(\frac1{2})=1$  and on each interval $(k, k+1)$ for $k\in \N^*$ with slope $g_\ell'(k+\frac1{2})=(h^*)'(k+\frac1{2}) = k+ 1$. The expression \eqref{fMexpli} for $f_\ell$ follows by computing $g_\ell^*$. 
  
As demonstrated above, optimal point configurations for a constant external potential are obtained by periodically grouping a suitable number of points. Therefore, optimal sets are essentially multisets (see Remark \ref{relax}).
 It is probable that a similar phenomenon occurs in higher dimensions. 
 
 \medskip
 The situation will vary if we consider a non-monotonic step function, such as the following:
$$\ell(r)=1 \ \text{for $r\in [0,1]$},\ \ell(r)=4 \ \text{for $r\in (1,2)$},\    
 \ell(r)=0\ \text{for $r\geq 2$}.$$ 
In this case, we anticipate that optimal configurations may be periodic, but associated with a non-uniform Voronoi tessellation, consisting of patterns of different sizes, as observed in the context of optimal location problem (see \cite[Sec 3.4]{BouJimRaj}).

\section{Proof of the main Theorem }

First, we check the properness property \eqref{epsupper} to make sure that the upper $\Gamma$- limit of $F_\e$ is not trivial.
To do this, it is sufficient to apply Lemma \ref{upperb-lemma} by choosing $u_0= a \,\one_\O$ for $a \in [0, r_0^{-d}]$. 
Next, by virtue of the equi-coercivity property of $F_\e$ proved in Proposition \ref{compactness-Feps} and 
of the Kuratowski compactness theorem (see Proposition \ref{essential} and the introductory comment of Subsection 3.2), 
we can find a sequence $\e_k\to 0$ and a weak* lower semicontinuous functional
$F: \MM_+(\O) \to [0,+\infty]$ such that $F_{\e_k} \Gconv F$ as $k\to\infty$.
Note that the limit $F$ may a priori depend on the chosen sequence $\e_k\to 0$. 
Accordingly, we will complete the proof of Theorem\ref{main-thm} in two steps which are outlined below:

\med {\bf Step 1:}\ we show that
$F$ is a local functional of the form
\begin{equation}\label{Ij}
F(\rho) :=  \begin{cases} \int_\O  j (x,u)\, dx & \text{if $\rho= u\, \LL^d \res \O$}\\
 +\infty & \text{otherwise} \end{cases}
\end{equation} 
 where  $j:\O\times  \R \to [0,+\infty]$ is  a suitable convex normal integrand such that $j(\cdot,0)=0$ a.e. in $\O$.
As a consequence $F$ is convex, weak* l.s.c. and coincides with its Fenchel biconjugate, i.e. :
 $$F(\rho)= F^{**}(\rho) = \sup_{v\in\CC(\Ob)} \left\{\int v\, d\rho - F^*(v) \right\}.$$

\med {\bf Step 2:}\ \  we identify the Fenchel conjugate  $F^*$ in terms of the convex function $g_\ell$ defined in\eqref{def:gell}, namely:
 $$ F^*(v) = \int_\O  g_\ell (v)\, dx \quad,\quad \text{for every $v\in \CC(\Ob)$}.   $$
 It follows that the limit functional $F$ \emph{does not depend on the sequence} $(\e_k)$. 
Also, since \eqref{superquadra} the convex function $f_\ell=g_\ell^*$ has a  superlinear growth at infinity,  by applying a classical result on convex functionals on measures (see for instance \cite{BouVal}), we obtain the equalities $F=F^{**}=F_\ell$ where:
$$ F_\ell(\rho) :=  \begin{cases} \int_\O  f_\ell (u)\, dx & \text{if $\rho= u\, \LL^d \res \O$}\\
 +\infty & \text{otherwise} \end{cases}.$$ 
 This will conclude the proof of the $\Gamma$-convergence of the whole sequence $(F_\e)$
 as stated in Theorem \ref{main-thm}.

 \med
 \begin{proof}[Proof of Step 1]  Let $\rho\in \MM_+(\Ob)$ be such that $F(\rho)<+\infty$. 
 Then, there exists a recovering sequence $\rho_k \weak \rho$ such that $\limsup_{k\to\infty} F_{\e_k}(\rho_k) = F(\rho)<+\infty$.
 By the assertion ii) of Proposition \ref{compactness-Feps}, we infer that $\rho$ is an absolutely continuous measure.
 Accordingly, there exists a functional  $J: L^1(\O) \to [0,+\infty]$ such that 
  $$ F(\rho) :=  \begin{cases} J(u) & \text{if $\rho= u\, \LL^d \res \O$}\\
 +\infty & \text{otherwise} \end{cases} .$$
The following result  will be crucial for deriving the integral representation and  the convexity of $J$. 
Its delicate proof is postponed to the end of this  section.

 \begin{lemma}\label{Jlist}  The functional $J: L^1(\O) \to [0,+\infty]$ defined above satisfies the following: 
 
 \begin{itemize}
%:
\item[(i)] \ $J$ is  weakly lower semicontinuous and satisfies $J(0)=0$;

\item [(ii)] \ The domain of $J$ is a subset of $L^1(\O;\R_+)$ and
$J( u\, \one_A) \le J(u)$  holds for every $u\in L^1(\O,\R_+)$ and every Borel subset $A\subset \O$;

\item [(iii)] \  It holds $J(u+v) =J(u)+ J(v)$ whenever $u\, v= 0$ is satisfied almost everywhere in $\O$.
\end{itemize}

 \end{lemma}
 
 In view of the assertions (i) and (iii) of Lemma \ref{Jlist} and since the Lebesgue measure on $\O$ is atomless, we may apply  a classical integral representation (see for instance Hiai and Umegaki  \cite{Hia,Umegaki} or
 the monograph \cite{Buttazzo-book}) according to which there exists a  suitable \emph{convex} normal integrand $j$ such that \eqref{Ij} holds. 
Moreover, as $J\ge 0$ and $J(0)=0$, we have $j(\cdot,0)=0$ a.e. while, due to the assertion (ii),  the integrand $j$ satisfies $j(x,t) =+\infty$ if $t<0$.  
\end{proof}

\med
 \begin{proof}[Proof of Step 2] \quad From Step 1 and by a classical result on  integral functionals (see for instance \cite{BouVal}), the Fenchel conjugate of $F$ 
 is given for every $v\in \CC(\Ob)$ by:
 $$ F^*(v) = \sup_{u\in L^1(\O)} \left\{ \int_\O v\, u\, dx - \int_\O j(x, u(x))\, dx\right\} \ =\  \int_\O j^*(x, v(x))\, dx .$$
 Obviously we may extend this equality to all  functions $v\in L^\infty(\O)$.
 Noticing that $j^*(x,0) = - \inf j(x,\cdot)= -j(x,0)=0$, we observe that, for every $\la\in\R$ and for every hypercube $Q(x_0,a) \subset \O$,
we have 
$$   F^*(\la \, \one_{Q(x_0,a)}) = \int_{Q(x_0,a)} j^*(x,\la)\, dx .$$
 Next we claim that, for any such an hypercube $Q(x_0,a) \subset \O$, the following holds:
\begin{equation}\label{claim=gell}
  F^*(\la \, \one_{Q(x_0,a)})\ =\ a^d\, g_\ell(\la)
\end{equation}
Suppose that this claim is true. Then, by considering Lebesgue points of $j^*(\cdot,\la)$ for  $\la$ in a dense countable subset $D$ of $\R$, we can find  a Lebesgue negligible subset $N\subset \O$ such that $j^*(x, \la) = g_\ell(\la)$  for all $(x,\la) \in (\O\setminus N)\times D$.
Thanks  to the continuity of $g_\ell$ proved in Proposition \ref{bounds-gell} and to the convexity of $j^*(x,\cdot)$,  the latter equality 
can be then extended to all   $(x,\la) \in (\O\setminus N)\times \R$, so that we have  $ F^*(v) = \int_\O  g_\ell (v)\, dx$
for every $v\in \CC(\Ob)$.  Hence the conclusion of Step 2 is reached and the proof of Theorem \ref{main-thm} is complete provided we can confirm \eqref{claim=gell}. 

\medskip
We now focus on the proof of  the equality \eqref{claim=gell}.
For $\la\le 0$, this equality is trivial since $g_\ell(\la)=j^*(x,\la)=0$.
Next we  observe that, for every $\la\ge 0$ and $Q(x_0,a) \subset \O$, we have:
\begin{align*}
 F_{\e_k}^*(\la \, \one_{Q(x_0,a)}) & := \sup \left\{ \la \, \rho(Q(x_0,a) - F_{\e_k}(\rho)\ :\ \rho\in \MM_+(\Ob)  \right\}\\
 &= \e_k^d  \sup_{S\subset \Ob}  \left\{\la \, \sharp(S\cap Q(x_0,a)) - \EE_{\e_k}(S)  \right\}  \\
 &= \e_k^d  \sup_{S\subset  Q(x_0,a)}  \left\{\la \, \sharp(S) - \EE_{\e_k}(S)  \right\}\\
 &= \e_k^d  \sup_{S'\subset  Q(x_0,\frac{a}{\e_k})}  \left\{\la \, \sharp(S') - \EE_\ell(S'/\e_k)  \right\} \\
 & = \e_k^d  \, \Gamma_\ell(\la,Q(x_0,\frac{a}{\e_k}))
\end{align*}
where:
\begin{enumerate}
\item [-] to pass from the second to the third line, we  substitute any competitor $S\subset\Ob$ with $S \cap Q(x_0,a)$ which has larger energy;
\item [-] to pass from the  third line to the two last  lines, we set $S= S'/\e_k$ for going back from the $\e_k$- scaled  energy \eqref{def:EEeps} to the ground interaction energy $\EE_\ell$ and ultimately recover the set function $\Gamma_\ell$ defined in \eqref{def:Gamma-ell}.  
\end{enumerate}
Therefore, thanks to \eqref{gellim}, we can pass to the  limit $k\to+\infty$ (the position of $x_0$ is irrelevant) and obtain the equality
$$  \lim_{k\to+\infty} F_{\e_k}^*(\la \, \one_{Q(x_0,a)})\ =\  a^d\, g_\ell(\la).$$
So, proving  \eqref{claim=gell} reduces to checking the equality
$$  \lim_{k\to+\infty} F_{\e_k}^*(\la \, \one_{Q(x_0,a)})  \ =\    F^*(\la \, \one_{Q(x_0,a)}),$$
 that we rewrite in the equivalent form:
\begin{equation}\label{cv-infima}
 \inf_{\rho\in \MM_+(\Ob)} \left\{ F_{\e_k}(\rho) - \la \, \rho(Q(x_0,a)) \right\} 
 \to  \inf_{\rho} \left\{  F(\rho) - \la \, \rho(Q(x_0,a)) \right\}. 
\end{equation}
The left hand side infimum in \eqref{cv-infima} being non-positive (easily seen by taking $\rho=0$ as a competitor), we may apply Proposition \ref{compactness-Feps}  with 
the choice $U=-\la\, \one_{Q(x_0,a)}$.  Therefore any minimizing sequence $(\rho_k)$ for the left hand side of \eqref{cv-infima} is bounded in $\MM_+(\Ob)$ hence weakly* relatively compact.
By the assertion iii) of  Proposition \ref{essential}, we will be able to conclude the convergence of infima in \eqref{cv-infima} if we can show that 
\begin{equation}\label{Gamma-sum}
F_{\e_k} + G\ \Gconv F+G \quad \text{being} \quad G(\rho):= - \la \, \rho(Q(x_0,a) .
\end{equation}
In virtue of Theorem \ref{main-thm}, we already know  that $F_{\e_k} \Gconv F$. Then it is enough to invoke the stability property of the assertion iv) of Proposition \ref{essential}.  However the functional $G$ given above  is not weak* continuous on $\M(\Ob)$ and therefore, we need to verify the less stringent requirements set out in \eqref{stability}. The first one is satisfied since $|G(\rho)|\le \la \, \|\rho\|$. For the second one, 
 we observe that, if $F(\rho)<+\infty$, then 
 $\rho$ is of the form $\rho=u\, \LL^d\res\O$ (see Proposition \ref{compactness-Feps}), hence  $\rho(\partial Q(x_0,a))=0$ and every sequence $\rho_n\weak \rho$ satisfies  $\rho_n(Q(x_0,a)) \to \rho(Q(x_0,a))=\int_{Q(x_0,a)} u\, dx$. 
 This confirms the  validity of \eqref{Gamma-sum}, hence that of \eqref{cv-infima}. As a result the equality \eqref{claim=gell} is proved
and, as announced, this achieves Step 2 and the proof of Theorem \ref{main-thm}.
 \end{proof}
%%%%%%%%%%%%%%%%%%%%%%
\medskip
\begin{proof}[Proof of Lemma \ref{Jlist}]
  (i)\  Let $u_k \to u$ in $L^1(\O)$. Then  $\rho_k= u_k\, \LL^d\res\O$ and $\rho= u\, \LL^d\res\O$ are such that
$\rho_k \weak \rho$ in $\MM_+(\Ob)$. Since $F=\Gamma-\lim F_{\e_k}$ is wzak* lower semicontinuous, we infer that
$$ \liminf_{k\to\infty} J(\u_k) = \liminf_{k\to\infty} F(\rho_k) \ge F(\rho) = J(u) .$$
To show that $J(0)=0$, we consider as $S_k$ a singleton $\{x_0\}$ so that the associated  measure $\rho_k= (\e_k)^d \, \delta_{x_0}$ satisfies $\rho_k \weak 0$ while $F_{\e_k}(\rho_k)=0$.

\med
(ii)\ By the definition of the $\Gamma$-limit,  $F(\rho)<+\infty$ implies that  $\rho$ is a weak* limit of a sequence $(\rho_k)$ in $\MM_+(\Ob)$, hence of the form $\rho=u\, \LL^d_\O$ with $u\ge 0$.  Let now $u\in L^1(\O;\R_+)$ and $A$ a Borel subset of $\O$.
We show first that $J(u\, \one_A)\le J(u)$ if $\LL^d(\partial A)=0$. We may assume that $J(u)<+\infty$. Hence, there exists a family of subsets $S_k\subset \O$
such that $\rho_k =\rho_{S_k}^{\e_k}$ satisfies $\rho_k\weak \rho:=u\, \LL^d\res\O$ and $F_{\e_k} (\rho_k) \to J(u)$. 
If we let $S_k'= S_k \cap A$ and $\rho_k'=\rho_{S_k'}^{\e_k}$, we have $\rho_k' \weak u\, \one_A \, \LL^d\res \O$, since indeed the convergence $\rho_k\weak \rho$ is tight while $\rho(\partial A)=0$. Therefore, $F_{\e_k}(\rho_k') \le F_{\e_k}(\rho_k) $. By passing to the limit, as $k\to+\infty$, we deduce that
$$ J(u\, \one_A) \le F(\rho\, \one_A) \le \liminf_{k\to\infty} F_{\e_k}(\rho_k') \le \limsup_{k\to\infty} F_{\e_k}(\rho_k)= J(u).$$
To extend the inequality to any Borel subset $A$ of $\Omega$, it is enough to consider an approximating sequence $(A_n)$ such that
$$  \LL^d(\partial A_n)= 0\quad,\quad \LL^d (A_n \Delta A) \to 0 ,$$
and then pass to the limit in the inequality $J(u)\ge J(u\one_{A_n})$ while letting $n\to+\infty$. Indeed, the conclusion will then follow from the lower semicontinuity of $J$
with respect to the norm convergence in $L^1(\O)$.
Now, to construct such a sequence $(A_n)$, we consider  a compact subset  $K_n \subset A$ and an open subset $\omega_n\supset A$ such that
$ \LL^d(\omega_n\setminus K_n)\le \frac1{n}.$ For every $n$, we can choose a suitable $r_n>0$ such that the enlarged open set
$A_n= K_n + B(0,r_n)$ satisfies $\LL^d(\partial A_n)=0$ \footnote{Here we use the fact that the function $\a_n(r)=\LL^d(\{x\in \O: {\rm dist}(x,K_n)>r\})$ is
bounded monotone non increasing so that it is continuous except possibly on a finite or countable subset of $\R_+$. } while $A_n\subset \omega_n$. Then clearly $ \LL^d (A_n \Delta A)\le  \LL^d (\omega_n\setminus K_n) \to 0$.

\medskip
Let us now prove now the assertion (iii). In a first step, we assume that $\spt(u)\cap \spt(v)=\emptyset$ so that there exists open subsets
 $A \supset \spt(u)$ and $B \supset \spt(v)$  such that ${\rm dist}(A,B):=  \eta >0.$
 
% step The condition $u\, v=0$ a.e.  means that there exists two Borel subsets $A, B\subset\O$ such that:
%%$A=\{u \not= 0\}$ and $B=\{v \not= 0\}$ up to a set of vanishing Lebesgue measure.  Therefore
% $$  A\cap B= \emptyset \quad,\quad u+v= \begin{cases}u & \text{a.e. in $A$} \\  v & \text{a.e. in $B$}
%   \end{cases}.$$

We begin by proving the inequality $J(u+v) \ge J(u)+ J(v)$.
 Without loss of generality, we may assume that $J(u+v)<+\infty$ (hence $u$ and $v$ are non-negative).  
 Then there exists a sequence of sets $S_k \subset \Ob$ such that
 $$  \rho_k = \mu_{S_k}^{\e_k} \weak (u+v) \, \LL^d\res\O \quad,\quad F_{\e_k} (\rho_k) \to J(u+v) .$$
 We write $S_k= S_k' \cup S_k''$ where $S_k'= S_k\cap A$ and $S_k''= S_k\cap B$ are disjoint. Then, we have
 $ \rho_k = \rho_k' +\rho_k''$  where $\rho_k' = \mu_{S_k'}^{\e_k}$ and $\rho_k'' = \mu_{S_k''}^{\e_k}.$
Clearly $\rho_k' \weak u\, \LL^d\res\O$ while $\rho_k'' \weak v\, \LL^d\res\O$. Therefore, by applying the $\Gamma-\liminf$ inequality to
$\rho_k'$ and $\rho_k''$ while taking into account the super-additivity property of $F_{\e_k}$ (see \eqref{subp-eps}), we deduce that:
\begin{align*} J(u+v) = \lim_{k\to\infty} F_{\e_k} (\rho_k) &\ge \liminf_{k\to\infty} F_{\e_k} (\rho_k') + \liminf_{k\to\infty} F_{\e_k} (\rho_k'') 
\\
&\ge F(u\, \LL^d\res\O) +F(v\, \LL^d\res\O) = J(u)+ J(v).
\end{align*}
To show the converse  inequality  $J(u+v) \le J(u)+ J(v)$, we assume without any loss of generality that $J(u)<+\infty$ and $J(v)<+\infty$. 
Then we consider recovering sequences $\rho_k' = \mu_{S_k'}^{\e_k}$ and $\rho_k'' = \mu_{S_k''}^{\e_k}$ such that $\rho_k' \weak u\, \LL^d\res\O\ , \rho_k'' \weak v\, \LL^d\res\O$ and 
$ F_{\e_k} (\rho_k)\to J(u)\ , F_{\e_k} (\rho_k'')\to J(v).$ 
Up to dropping the elements of $S_k'$ which ar not in $A$ and the elements of $S_k''$ which are not in $B$, we may assume that
$\spt(\rho_k')\subset A$ and $\spt(\rho_k')\subset B$. Indeed, removing these points will not affect the weak* convergence to $u\, \LL^d\res\O$ and
$v\, \LL^d\res\O$ respectively while the total energy $\EE_{\e_k}$ will not increase. Therefore, by exploiting the right hand inequality in 
\eqref{subp-eps} and since  $ \rho_k = \rho_k' +\rho_k''$ converge weakly* to $\rho=(u+v) \, \LL^d\res\O$, we are led to the following set of inequalities:
\begin{align*} J(u+v) \le \liminf_{k\to\infty} F_{\e_k} (\rho_k ) &\le \limsup_{k\to\infty} F_{\e_k} (\rho_k') + \limsup_{k\to\infty} F_{\e_k} (\rho_k'')\\ & +  \limsup_{k\to\infty} \frac{ 2 \ell_+ (\eta/\e_k) }{\e_k^d}  \  \|u\|_{L^1(\O)}  \|v\|_{L^1(\O)} 
\\
&\le J(u) + J(v)  + C \, \limsup_{k\to\infty} \frac{\ell_+ (\eta/\e_k) }{\e_k^d}\ , \end{align*}
where, in the second line, we used the tight convergence of $\rho_k', \rho_k''$. 
The conclusion follows by noticing that $\ell_+$ coincides with $\ell$ for large values where it is non-increasing. Thus the integrability condition (H3) implies that $r^d \, \ell(r) \to 0$ as $r\to+\infty$. The desired sub-additivity inequality follows.

\medskip
In a second step,  we remove the strict separation condition on the supports of $u$ and $v$ by simply assuming that the upper-level sets $A:= \{u>0\}$ and $B:= \{v>0\}$ satisfy $\LL^d(A\cap B)=0$ (which, for $u,v$ non-negative, is equivalent to say that  $u\, v=0$ a.e.).
To that aim, possibly after substituting $A,B$ with non-intersecting  Borel representatives, we consider increasing sequences of compact subsets $K_n'\subset A, K_n''\subset B$ such that:
 $$\LL^d(A\setminus K_n')\to 0\ ,\ \LL^d(B\setminus K_n'')\to 0\ ,\  K_n' \cap K_n'' =\emptyset. $$
Then $u_n:= u \, \one_{K'_n}, v_n:= v \, \one_{K''_n}$ satisfy $J(u_n+v_n)=J(u_n)+J(v_n).$  
In virtue of assertion (ii), we infer that, for every $n$:
$$  J(u_n+v_n) \le J(u) + J(v) \quad\mbox{ and} \quad  
J(u_n) + J(v_n)\le J(u+v) .$$
 In virtue of the lower semicontinuity of $J$, since $u_n\to u$ and $v_n\to v$ in $L^1(\O)$, we deduce from above  the inequalities $J(u+v)\le J(u) + J(v)$ and  $J(u) + J(v)\le J(u+v)$, hence the desired additivity property.

\end{proof}

\bibliographystyle{plain}

\end{document}